\documentclass[journal]{IEEEtran}

\usepackage{amsmath,amsfonts, amssymb,epsfig,epstopdf,bm}
\usepackage{caption,lipsum,rotating,makecell,stfloats, eufrak, setspace,cite}
\usepackage{graphicx,multirow,multicol,threeparttable}
\graphicspath{{./}{figures/}}
\usepackage{diagbox}
\usepackage{eufrak}
\usepackage[mathscr]{eucal}
\usepackage[table]{xcolor}
\definecolor{mygray}{gray}{0.8}
\usepackage{url}
\usepackage{algorithm,algorithmic}
\usepackage{amsthm}
\def\mF{f}
\def\mG{g}
\def\mE{\mathcal{E}}
\def\mL{\mathcal{L}}
\def\mS{\mathcal{S}}

\def\mT{\mathcal{T}}

\theoremstyle{plain}

\newtheorem{defn}{Definition}

\theoremstyle{remark}

\begin{document}

\title{Nonlinear Chaotic Processing  Model}

\author{Zhongyun Hua,~\IEEEmembership{Student Member,~IEEE,}
        Yicong~Zhou,~\IEEEmembership{Senior Member,~IEEE}
\thanks{This work was supported in part by the Macau Science and Technology Development Fund under Grant FDCT/016/2015/A1 and by the Research Committee at University of Macau under Grants MYRG2014-00003-FST and MYRG2016-00123-FST.}
\thanks{Z. Hua is with School of Computer Science and Technology, Harbin Institute of Technology Shenzhen Graduate School, Shenzhen 518055, China, and also with Department of Computer and Information Science, University of Macau, Macau, China (e-mail: huazyum@gmail.com).}
\thanks{Y.~Zhou is with Department of Computer and Information Science, University of Macau, Macau, China (e-mail: yicongzhou@umac.mo).}
}

\iffalse
\markboth{A manuscript submitted to IEEE transaction on circuits and systems,~2016}%
{Shell \MakeLowercase{\textit{et al.}}: Bare Demo of IEEEtran.cls for Journals}
\fi
\maketitle

\begin{abstract}
Designing chaotic maps with complex dynamics is a challenging topic. This paper introduces the nonlinear chaotic processing (NCP) model, which contains six basic nonlinear operations. Each operation is a general framework that can use existing chaotic maps as seed maps to generate a huge number of new chaotic maps. The proposed NCP model can be easily extended by introducing new nonlinear operations or by arbitrarily combining existing ones. The properties and chaotic behaviors of the NCP model are investigated. To show its effectiveness and usability, as examples, we provide four new chaotic maps generated by the NCP model and evaluate their chaotic performance using Lyapunov exponent, Shannon entropy, correlation dimension and initial state sensitivity. The experimental results show that these chaotic maps have more complex chaotic behaviors than existing ones.
\end{abstract}


\section{Introduction}%
\IEEEPARstart{C}HAOTIC behaviors widely exist in many natural and non-natural phenomena, such as the weather and climate~\cite{Ivancevic2008SSBM}. It can be studied through some analytical techniques or mathematical models, known as chaotic systems. Although there is no universally accepted mathematical definition for chaos, a chaotic system with chaotic behavior always displays the following properties: initial state sensitivity, topological transitivity and density of periodic orbits~\cite{Boris2003first}. Thus, the future behavior of a chaotic system is fully determined by its initial state. Any arbitrarily tiny change in the initial state results in a totally different orbit. With these significant properties, chaos theory has wide applications in different fields of science and engineering~\cite{stockmann2006quantum,Strogatz2001PP,Shen2014IECS1}, especially in cryptography and communication~\cite{Millerioux2008IETCS1,Cho2015IETCS1,Chou2013IETIFS,Dimassi2011IETCS1}. This is due to the facts that many properties of chaotic behavior can be found similar counterparts in cryptography and the synchronization of chaos is extremely suitable for designing secure communication systems~\cite{Habutsu_EUROCRYPT91,Patidar2009IS,Ashwin2003Nature}.

A dynamical system with chaotic attractors is globally stable but locally unstable. This means that arbitrarily close states diverge from each other but never depart from the attractor. However, the phase plane of finite precision platforms cannot have infinite number of states. When chaotic behavior is simulated in a finite precision platform, the arbitrarily close states will overlap, and thus the chaotic behavior will degrade to periodic behavior~\cite{Li2009IVC,Wong2010IETCS2,Zhu2008JSEE}. If states of a chaotic attractor are more concentrated, the extremely close ones are more possible to overlap. Thus, a chaotic system with good ergodicity is desired in real applications. On the other hand, with the development of discerning chaos technologies, some chaotic systems with simple definitions and behaviors can be easily attacked using different methods~\cite{Srivastava2009IETSMCC}. Recently, many studies are performed to predict chaotic behaviors by estimating their states~\cite{Liu2012IETSMC}, identifying their chaotic signals~\cite{Ling1999SP,Zhu2002IETCS1}, or deducing their initial conditions~\cite{Wu2004CSF,Lin2012IETSP}. If the future behavior of a chaotic system is successfully predicted, its corresponding chaos-based applications may have the high probability of being broken~\cite{Skrobek2007PL,Yang1998PL}.

Recently, a wide body of research has devoted to developing new dynamical systems with complex behaviors. These studies can be classified into two catalogs: designing specific chaotic maps and developing methodologies of generating a series of chaotic maps.
The former aims to produce well-defined chaotic maps with clear mathematical definitions, such as the L{\"u} attractor~\cite{Lu2002IJBC}, the multiwing chaotic attractors~\cite{Yu2011IETCS2,Huang2015IETCS2,Chen2008CSF} and the multiscroll chaotic attractors~\cite{Lu2005IETCS1,Zuo2014IETCS2}. The latter is to propose a framework or a system that can generate a series of chaotic maps, such as the cascade chaotic system~\cite{Zhou2015IETC}, the wheel-switching chaotic system~\cite{Wu2014IETCS1}, the coupling scheme~\cite{Wang2013IETAC,Pagliari2011IETMC,Nkomo2013PRL} and the parameter-control chaotic system~\cite{Hua2016IETC}.

Motivated by the arithmetic operations in the field of real number, this paper proposes a nonlinear chaotic processing (NCP) model for generating chaotic maps. The NCP model includes six basic nonlinear operations, i.e. the cascade, modulation, switching, fusion, scalar cascade and scalar modulation. Each operation is a general framework and able to use existing chaotic maps as seed maps to generate new chaotic maps. Compared with seed maps, the generated chaotic maps usually have more complex chaotic behaviors. In addition to these six basic operations, we can include other nonlinear operations or arbitrarily combine the existing ones to extend the NCP mode. Properties of the NCP model are comprehensively analyzed and its chaotic behavior is investigated using Lyapunov exponent (LE). Using three existing 1D chaotic maps as seed maps, four examples of new chaotic maps are generated by the NCP model. We investigate their dynamical properties including equilibrium points, stability and bifurcation diagrams. Their chaotic performance is evaluated using LE, Shannon entropy (SE), correlation dimension (CD) and initial state sensitivity. Experimental results show that these four chaotic maps have wider chaotic ranges, better ergodicity and more complex chaotic behaviors than existing ones.

The rest of this paper is organized as follows. Section~\ref{section2} reviews the properties of three existing 1D chaotic maps as background. Section~\ref{section3} introduces the NCP model and its chaotic behavior is discussed in Section~\ref{section4}. Section~\ref{Section5} presents four examples of new chaotic maps generated by the NCP model and their performance is evaluated in Section~\ref{section6}. Section~\ref{section7} concludes this paper.

\section{Background}
\label{section2}
This section briefly reviews three widely used first-order difference equations, i.e. 1D chaotic maps, and analyzes their dynamics properties. They will be used as seed maps to demonstrate the NCP model in Section~\ref{Section5}.

The Logistic map is a first-order difference equation that arises in many contexts in the economic, social and biological sciences~\cite{May1976Nature}. It is represented as
\begin{equation*}
  x_{i+1} = \mL(x_i) = 4px_i(1-x_i).
\end{equation*}
When its parameter $p\in[0,1]$ and variable $x_i$ is limited into the interval $[0,1]$, it shows non-trivial dynamical behavior.

\begin{figure}[htbp]
\centering
\begin{minipage}[b]{0.99\linewidth}
  \begin{minipage}[b]{.32\linewidth}
    \centerline{\includegraphics[width=1\linewidth]{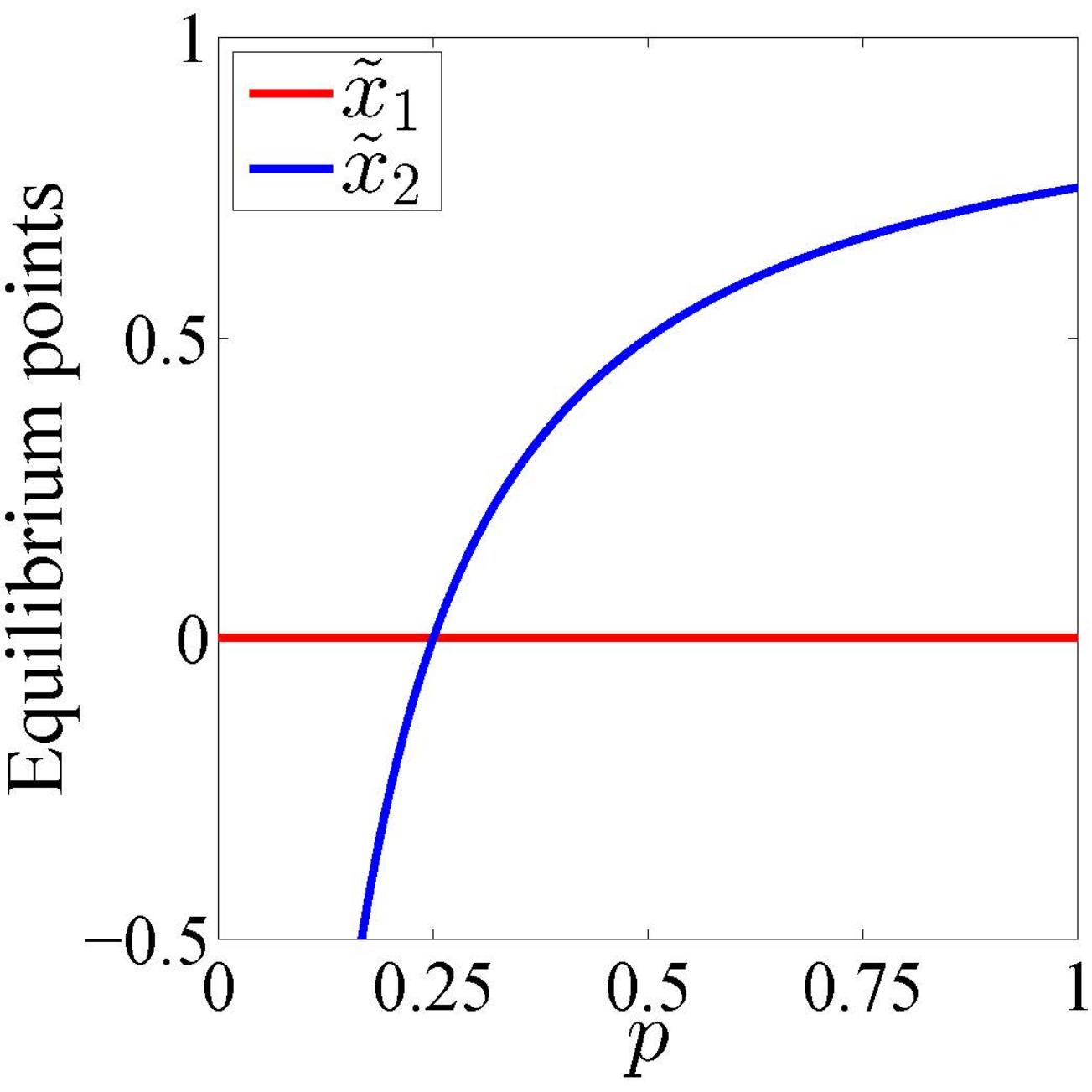}}
    \vspace{3pt}
    \centerline{\includegraphics[width=1\linewidth]{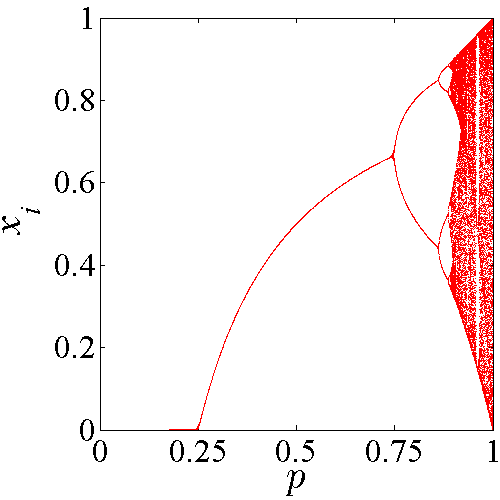}}
    \centerline{(a)}
\end{minipage}\hfill
  \begin{minipage}[b]{.32\linewidth}
    \centerline{\includegraphics[width=1\linewidth]{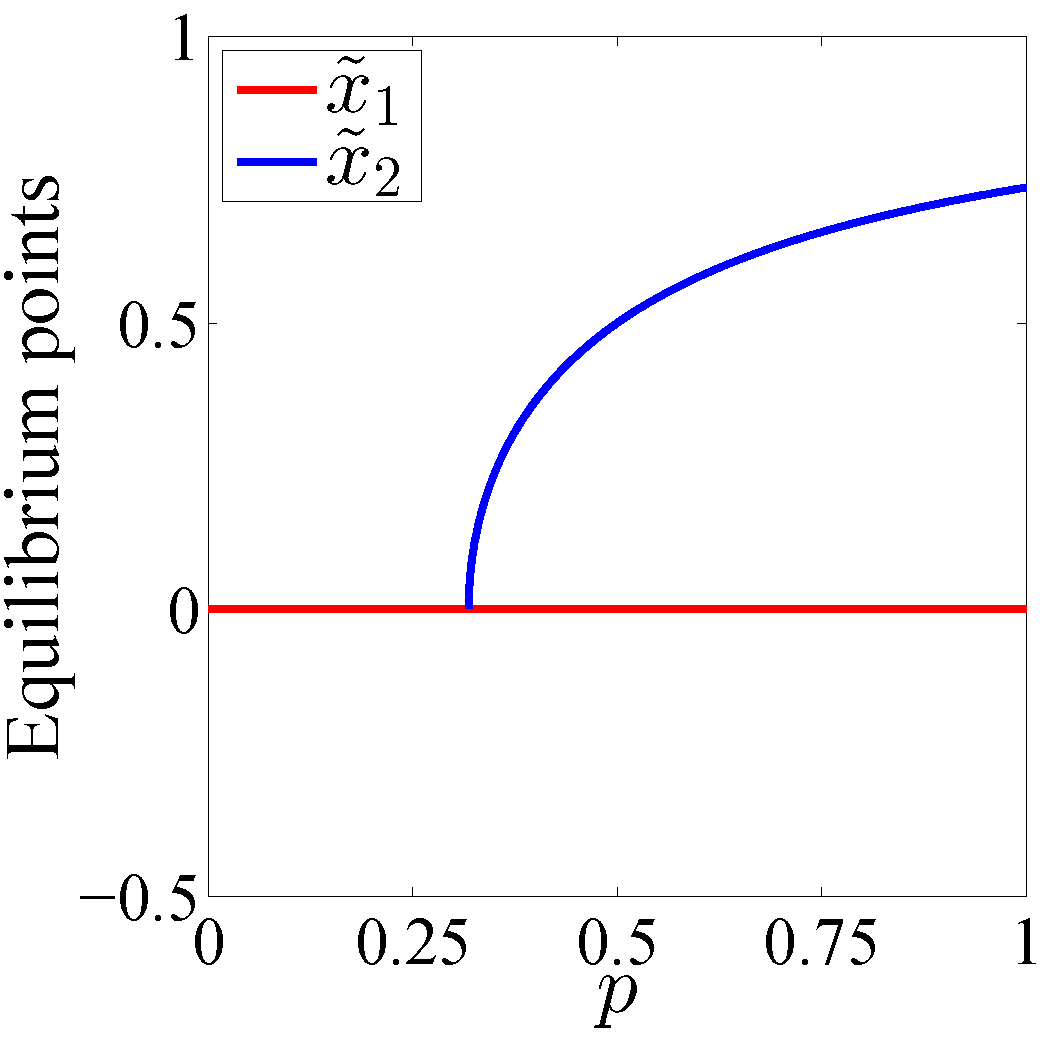}}
    \vspace{3pt}
    \centerline{\includegraphics[width=1\linewidth]{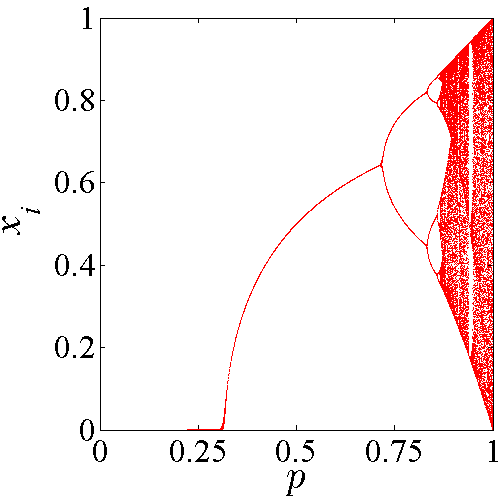}}
    \centerline{(b)}
\end{minipage}\hfill
  \begin{minipage}[b]{.32\linewidth}
    \centerline{\includegraphics[width=1\linewidth]{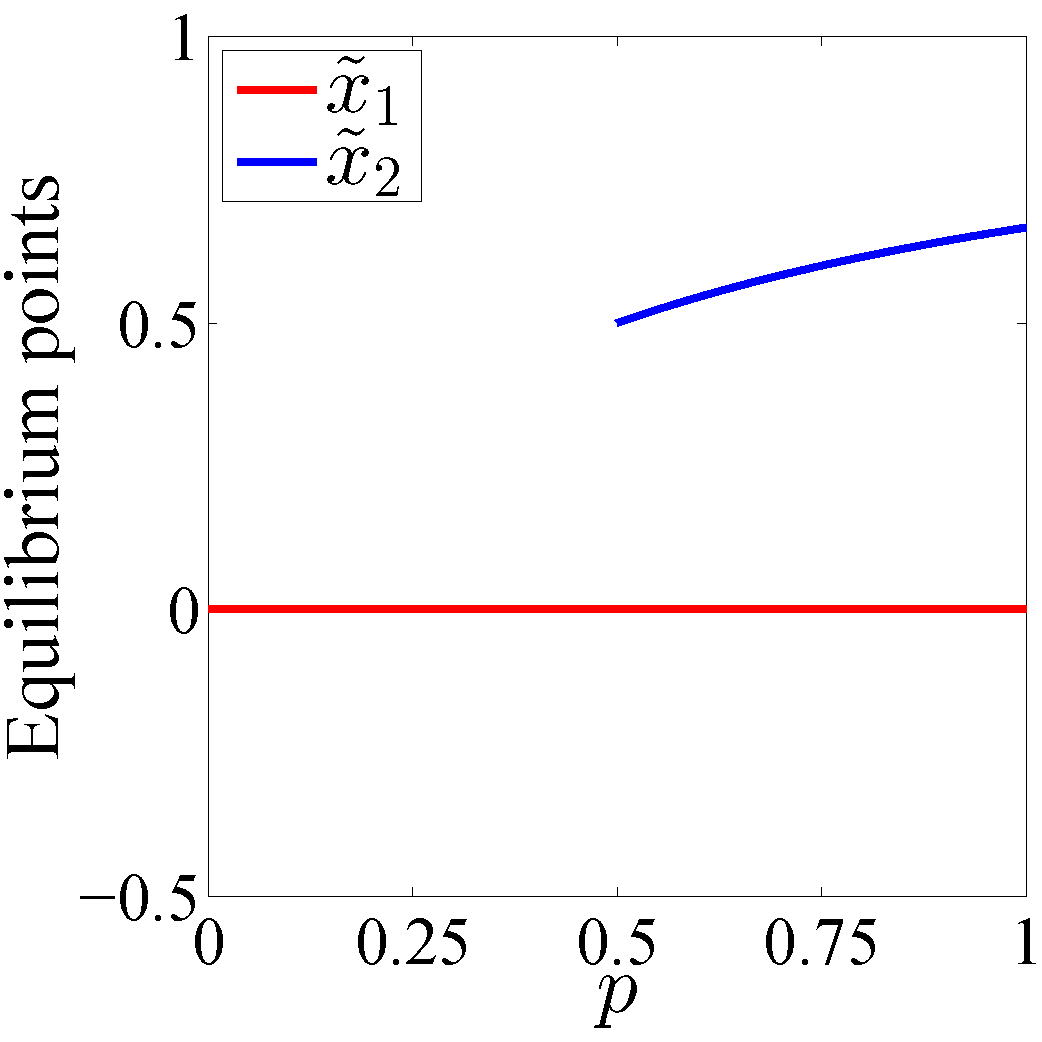}}
    \vspace{3pt}
    \centerline{\includegraphics[width=1\linewidth]{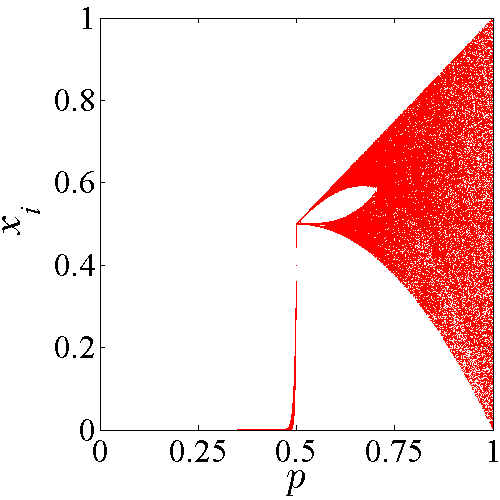}}
    \centerline{(c)}
\end{minipage}\hfill
\end{minipage}\hfill
    \caption{Equilibrium points and bifurcation diagrams of the (a) Logistic map, (b) Sine map and (c) Tent map.}
    \label{Fig.ExistingMaps}
\end{figure}

Equilibrium point (or fixed point) is the element of a function's domain that maps to itself. For a function $x_{i+1}=\mF(x_i)$, a graphical method of finding its equilibrium points is to find the points where its curve intersects the $45^\circ$ line, namely, $x_{i+1}=x_i$. Thus, the Logistic map's equilibrium point $\tilde{x}$ satisfies the equation
\begin{equation}\label{eq.fixLo}
  \tilde{x} = 4p\tilde{x}(1-\tilde{x}).
\end{equation}
Solve Eq.~\eqref{eq.fixLo}, we can get that the Logistic map has two equilibrium points, namely, $\tilde{x}_1=0$ and $\tilde{x}_2=1-1/(4p)$. The bifurcation diagram is to straightforwardly display the asymptotically approached or visited output values of a dynamical system with different parameter settings. Fig.~\ref{Fig.ExistingMaps}(a) plots the two equilibrium points and bifurcation diagram of the Logistic map with the change of its parameter $p$. Observed from the bifurcation diagram, we can get that the Logistic is chaotic when $p\in[0.9,1]$.

The Sine map is derived from the Sine function that maps the input angle within interval $[0,1]$ into the same interval. Mathematically, the Sine map is defined as
\begin{equation*}
  x_{i+1} =\mS(x_i) = p\sin(\pi x_i),
\end{equation*}
where the control parameter $p\in[0,1]$. To find out the Sine map's equilibrium point $\tilde{x}$, we set
\begin{equation}\label{eq.fixSine}
  \tilde{x}=p\sin(\pi \tilde{x}).
\end{equation}
Solve Eq.~\eqref{eq.fixSine}, we obtain that the Sine map has equilibrium point $\tilde{x}_1=0$ in the whole parameter range and another equilibrium point $\tilde{x}_2$ when its parameter $p>0.3184$. Fig.~\ref{Fig.ExistingMaps}(b) shows its equilibrium points and bifurcation diagram. The Sine map has chaotic behavior when $p\in[0.87,1]$.

The Tent map is a piecewise function that either scales or folds the input value based on its range. Mathematically, its generalized form can be defined as
\begin{equation*}
  x_{i+1} = \mT(x_i)=\begin{cases}
    2px_i, \ &\mbox{for}\ x_i<0.5,\\
    2p(1-x_i),\ &\mbox{for}\ x_i\geq 0.5,
  \end{cases}
\end{equation*}
where the parameter $p\in[0,1]$. To find out its equilibrium point $\tilde{x}$, we set
\begin{equation}\label{eq.fixTent}
  \tilde{x} = 2p\min\{\tilde{x},1-\tilde{x}\}.
\end{equation}
From Eq.~\eqref{eq.fixTent}, we calculate out that the Tent map has equilibrium point $\tilde{x}_1=0$ in the whole parameter range and equilibrium point $\tilde{x}_2=(2p)/(2p+1)$ in the range $p\in[0.5,1]$. The two equilibrium points and bifurcation diagram of the Tent map are plotted in Fig.~\ref{Fig.ExistingMaps}(c). The Tent map has chaotic behavior when $p\in(0.5,1)$.

\section{The NCP Model}
\label{section3}
This section introduces the nonlinear chaotic processing (NCP) model and its properties.

\begin{table*}[htbp]
\newcommand{\tabincell}[2]{\begin{tabular}{@{}#1@{}}#2\end{tabular}}
\renewcommand{\arraystretch}{1.4}
\setlength{\tabcolsep}{10pt}
\begin{center}
\caption{Definitions of six basic nonlinear operations in the NCP model.}
\label{tab.Operations}
\begin{tabular}{|l|c|c|} \hline

    Operations & Descriptions & Definitions \\\hline\hline
      Cascade ($\circledcirc$)        & $\mathcal{C}(x) = \mF(x)\circledcirc\mG(x)$ & $x_{i+1}  =\mG(\mF(x_i))$ \\\hline
  Modulation ($\odot$)       & $\mathcal{M}(x) = \mF(x)\odot\mG(x)$ &     \tabincell{l}{ $x_{i+1} =  \mG(r_{i+1},x_i)$, where  \\  $r_{i+1}=R(y_{i+1})$, $y_{i+1} =  \mF(y_i)$ } \\\hline

  Switching ($\circledast$)       & $\mathcal{W}(x) = \mF_1(x)\circledast\mF_2(x)\circledast\cdots \circledast\mF_l(x)$ &
                            \tabincell{l}{ $x_{i+1} = \mF_{q_i}(x_i) $, where \\ $q_i\in\{1,2,\cdots,l\}$}

  \\\hline
  Fusion ($\oplus$)     & $\mathcal{P}(x) = \mF(x)\oplus \mG(x)$ & $x_{i+1}  = (\mF(x_i) + \mG(x_i)) \mod 1$ \\\hline
Scalar cascade($\widetilde{\circledcirc}$) &  $\mathcal{U}(x)=c\ \widetilde{\circledcirc}\ \mF(x)$ & $\mathcal{U}(x) = \underbrace{\mF(x)\circledcirc\mF(x)\circledcirc \cdots\circledcirc\mF(x)}_c$ \\\hline
Scalar modulation ($\widetilde{\odot}$)& $\mathcal{D}(x) = c \ \widetilde{\odot}\ \mF(x)$ & $\mathcal{D}(x) = \underbrace{\mF(x)\odot\mF(x)\odot \cdots\odot\mF(x)}_c$ \\\hline

\end{tabular}
\end{center}
\end{table*}
\subsection{The NCP Model}
The proposed NCP model is shown in Table~\ref{tab.Operations}. It contains, but not limited to, six basic nonlinear operations. Each operation can use existing chaotic maps as seed maps to generate new ones. Next, we discuss these operations one by one.

\subsubsection{Cascade}
\begin{figure}[htbp]
\centering
\scriptsize
    \centerline{\includegraphics[width=0.65\linewidth]{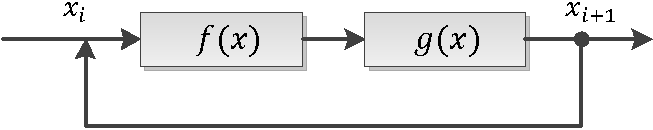}}
\caption{The cascade operation.}\label{fig.cascade}
\end{figure}

Motivated by the cascade operation in circuit design, we proposed a cascade operation for generating new chaotic maps in~\cite{Zhou2015IETC}. It connects two 1D chaotic maps in series. The definition of the cascade operation is given in Table~\ref{tab.Operations} and its structure is shown in Fig.~\ref{fig.cascade}. As can be seen, $\mF(x)$ and $\mG(x)$ are two 1D chaotic maps that are used as seed maps. The output of $\mF(x)$ is fed into the input of $\mG(x)$, and the output of $\mG(x)$ is the iterative value, and also feed back into the input of $\mF(x)$ for next iteration. The cascade operation has properties as follows:
\begin{itemize}
  \item  Non-commutativity: Exchanging the order of $\mF(x)$ and $\mG(x)$ will result in a different chaotic map, i.e. $\mF(x)\circledcirc\mG(x)$ and $\mG(x)\circledcirc\mF(x)$ are two completely different chaotic maps.
  \item  The cascade operation usually results in a chaotic map with more complex behaviors than its two seed maps, because its definition contains the mathematical concepts of its two seed maps.
\end{itemize}

\subsubsection{Modulation}
\begin{figure}[htbp]
\centering
\scriptsize
\begin{minipage}[b]{0.55\linewidth}
    \centerline{\includegraphics[width=1\linewidth]{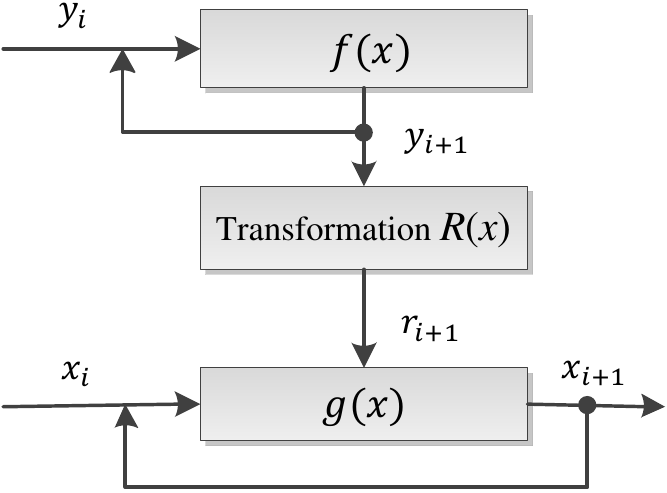}}
\end{minipage}\hfill
    \caption{The modulation operation.}\label{fig.modulation}
\end{figure}

The modulation operation uses the output of a chaotic map to dynamically control the parameter of another chaotic map to exhibit chaotic behaviors~\cite{Hua2016IETC}. Its structure is shown in Fig.~\ref{fig.modulation} and its definition is shown in Table~\ref{tab.Operations}, in which $\mF(x)$ and $\mG(x)$ are two 1D chaotic maps, the transformation $R(x)$ is to linearly transform the output of $\mF(x)$ into $\mG(x)$'s chaotic range, and $\mG(x)$ uses the dynamically changed parameter to generate trajectories.
The modulation operation has following properties:
\begin{itemize}
  \item Non-commutativity: Exchanging two seed maps, the modulation operation generates a completely different chaotic map, i.e. $\mF(x)\odot\mG(x)$ and $\mG(x)\odot\mF(x)$ are two totally different chaotic maps.
  \item The chaotic attractor of the modulation operation has a higher degree of freedom than those of its seed maps, because it can achieve a dynamically changed parameter in each iteration.
\end{itemize}

\subsubsection{Switching}
\begin{figure}[htbp]
\centering
\scriptsize
    \centerline{\includegraphics[width=0.7\linewidth]{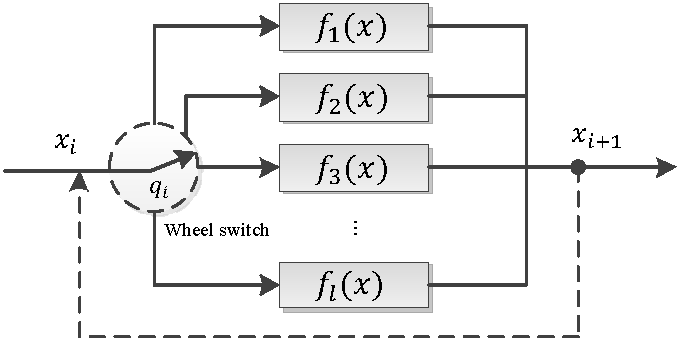}}
    \caption{The switching operation.}\label{fig.switching}
\end{figure}

The switching operation utilizes a wheel switch to select one of seed maps to execute in each iteration~\cite{Wu2014IETCS1}. As shown from its structure in Fig.~\ref{fig.switching}, the switching operation contains $l$ 1D normalized chaotic maps as seed maps and a controlling wheel switch $\mathbf{q}$. According to the pre-defined rules in $\mathbf{q}$, one seed map is selected to generate chaotic orbit in each iteration. The definition of the switching operation is shown in Table~\ref{tab.Operations}. As can be seen, $\mF_1(x), \mF_2(x), \cdots, \mF_l(x)$ are $l$ normalized chaotic maps and $\mathbf{q}=\{q_1,q_2,\cdots,q_l\}$ is the wheel switch, in which $q_i\in\{1,2,\cdots,l\}$. In the $i$-th iteration, the $q_i$-th seed map $\mF_{q_i}(x)$ is selected to execute.

\subsubsection{Fusion}
\begin{figure}[htbp]
\centering
\scriptsize
    \centerline{\includegraphics[width=0.7\linewidth]{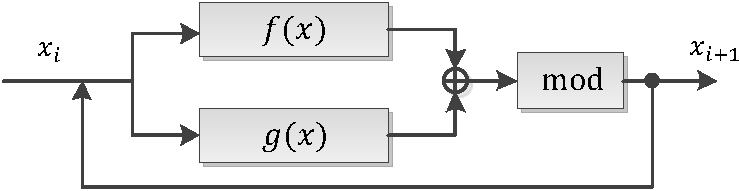}}
    \caption{The fusion operation.}\label{fig.fusion}
\end{figure}

The fusion operation generates new chaotic maps by mixing the dynamics of two seed maps in a nonlinear way~\cite{Zhou2014SP}. Its definition is shown in Table~\ref{tab.Operations} and structure is displayed in Fig.~\ref{fig.fusion}. In each iteration, the input is concurrently fed into two seed maps, and then the outputs of the two seed maps are combined by the modular arithmetic. Because the input is fed into two seed maps simultaneously and their outputs are added together, the fusion operation has the property of commutativity. Exchanging the positions of two seed maps, the generated chaotic maps are the same, i.e. $\mF(x)\oplus \mG(x)=\mG(x)\oplus \mF(x)$.

\subsubsection{Scalar Cascade}
\begin{figure}[htbp]
\centering
\scriptsize
    \centerline{\includegraphics[width=0.8\linewidth]{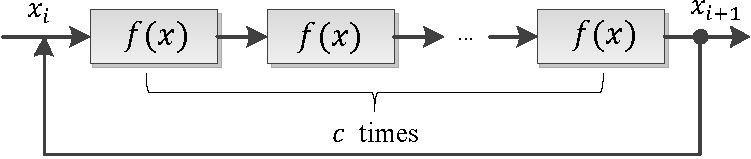}}
    \caption{The scalar cascade operation.}\label{fig.Mul}
\end{figure}

The scalar cascade operation generates chaotic maps by cascading a chaotic map with itself several times. Its definition is shown in Table~\ref{tab.Operations} and its structure is demonstrated in Fig.~\ref{fig.Mul}. The integer $c$ indicates how many times the seed map $\mF(x)$ is cascaded with itself. The scalar cascade operation has all the properties of the cascade operation.

\subsubsection{Scalar Modulation}
\begin{figure}[htbp]
\centering
\scriptsize
    \centerline{\includegraphics[width=0.55\linewidth]{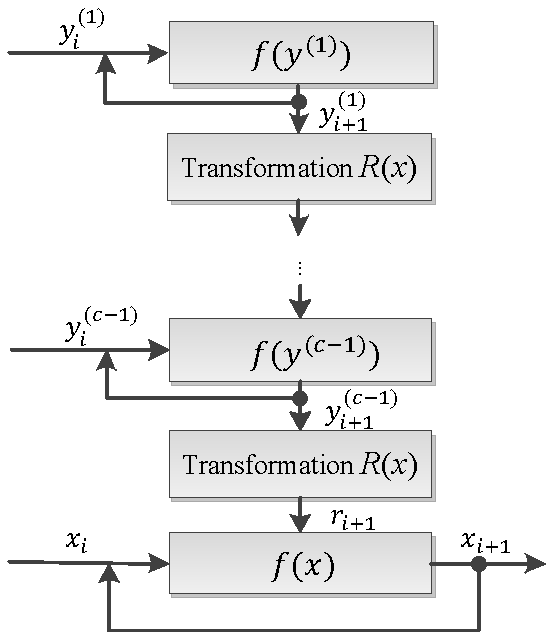}}
    \caption{The scalar modulation operation.}\label{fig.ScMo}
\end{figure}

The scalar modulation is defined in Table~\ref{tab.Operations} and its structure is shown in Fig.~\ref{fig.ScMo}. The $c$ is an integer and $c> 2$. The outputs of $s$-th $(1\leq s\leq c-2)$ control map $\mF(y^{(s)})$ are used to dynamically control the parameter of the $(s+1)$-th control map, and the outputs of the $(c-1)$-th control map $\mF(y^{(c-1)})$ are used to dynamically control the parameter of the seed map $\mF(x)$ to generate iterative values. The transformation $R(x)$ is to map the output of the $s$-th $(1\leq s\leq c-1)$ control map $\mF(y^{(s)})$ into the chaotic range of the $(s+1)$-th control map $\mF(y^{(s+1)})$ or seed map $\mF(x)$.

%\begin{table*}[htbp]
%\newcommand{\tabincell}[2]{\begin{tabular}{@{}#1@{}}#2\end{tabular}}
%\renewcommand{\arraystretch}{1.5}
%\setlength{\tabcolsep}{12pt}
%\begin{center}
%\caption{Definitions of nigh new chaotic maps.}
%\label{tab.Eg}
%\begin{tabular}{|c|l|} \hline
%
%  Operations    & Definitions \\\hline\hline
%         $ \mF_1(x)\oplus \mF_2(x)\circledcirc\mF_3(x)$ &     $x_{i+1} = \mF_3((\mF_1(x_i)+\mF_2(x_i)) \mod1 )$ \\\hline
%   $\mF_2(x)\oplus \mF_3(x)\circledcirc\mF_1(x)$ &     $x_{i+1} = \mF_1((\mF_2(x_i)+\mF_3(x_i)) \mod1 )$ \\\hline
%   $\mF_3(x)\oplus \mF_1(x)\circledcirc\mF_2(x)$ &     $x_{i+1} = \mF_2((\mF_3(x_i)+\mF_1(x_i)) \mod1 )$ \\\hline
%   $ \mF_1(x)\circledcirc \mF_2(x)\oplus \mF_3(x)$ &     $x_{i+1} = (\mF_2(\mF_1(x_i)) + \mF_3(x_i))\mod 1 $  \\\hline
%   $\mF_2(x)\circledcirc \mF_1(x)\oplus \mF_3(x)$ &     $x_{i+1} = (\mF_1(\mF_2(x_i)) + \mF_3(x_i))\mod 1 $  \\\hline
%   $\mF_2(x)\circledcirc \mF_3(x)\oplus \mF_1(x)$ &     $x_{i+1} = (\mF_3(\mF_2(x_i)) + \mF_1(x_i))\mod 1 $  \\\hline
%   $ \mF_3(x)\circledcirc \mF_2(x)\oplus \mF_1(x)$ &     $x_{i+1} = (\mF_2(\mF_3(x_i)) + \mF_1(x_i))\mod 1 $  \\\hline
%   $\mF_3(x)\circledcirc \mF_1(x)\oplus \mF_2(x)$    &     $x_{i+1} = (\mF_1(\mF_3(x_i)) + \mF_2(x_i))\mod 1 $  \\\hline
% $ \mF_1(x)\circledcirc \mF_3(x)\oplus \mF_2(x)$    &     $x_{i+1} = (\mF_3(\mF_1(x_i)) + \mF_2(x_i))\mod 1 $  \\\hline
%\end{tabular}
%\end{center}
%\end{table*}
\subsection{Extension of the NCP model}
The proposed NCP model is an opened model that is specially designed for chaotic systems. Except for the above six basic nonlinear operations presented in Table I, the users have flexibility to introduce new operations to extend the NCP model. On the other hand, the six basic nonlinear operations can be arbitrarily combined to form new operations. This is another straightforward way to further extend the NCP model. For $N$ seed maps $\mF_1(x)$, $\mF_2(x)$, $\cdots$, $\mF_N(x)$, a combination of basic operations can be defined as
\begin{equation}\label{eq.co}
  \mathcal{CO}(x) = \mF_1(x)\ \maltese\ \mF_2(x)\ \maltese \cdots\ \maltese\ \mF_N(x),
\end{equation}
where $\maltese$ represents one of the six basic operations listed in Table~\ref{tab.Operations}. Note that if $\maltese$ represents the scalar cascade or scalar modulation, $\mF_i(x)$ on the left of $\maltese$ is an integer constant instead of a chaotic map.

In Eq.~\eqref{eq.co}, any different or same chaotic maps and basic operations can be arbitrarily selected for combination to generate a large number of new chaotic maps. Table~\ref{tab.Eg} provides examples of combination when $N=3$. The two basic operations are set as the cascade operation $\circledcirc$ and modulation operation $\odot$. A total number of 12 types of operations can be generated and their definitions are displayed in Table~\ref{tab.Eg}. In these examples, the three seed maps, $\mF_1(x)$, $\mF_2(x)$ and $\mF_3(x)$, can be arbitrarily selected as different or same chaotic maps.

\begin{table}[htbp]
\newcommand{\tabincell}[2]{\begin{tabular}{@{}#1@{}}#2\end{tabular}}
\renewcommand{\arraystretch}{1.5}
\setlength{\tabcolsep}{2.8pt}
\begin{center}
\caption{Twelve combination examples of basic operations.}
\label{tab.Eg}
\begin{tabular}{|l|l|} \hline

  Operations    & Definitions \\\hline\hline
        $\mF_1(x)\odot \mF_2(x) \circledcirc \mF_3(x)$ &     \tabincell{l}{ $x_{i+1} =  f_3(f_2(r_{i+1},x_i))$, where  \\  $r_{i+1}=R(y_{i+1})$, $y_{i+1} =  f_1(y_i)$ } \\\hline
         $ \mF_1(x)\odot \mF_3(x) \circledcirc \mF_2(x)$ &     \tabincell{l}{ $x_{i+1} =  f_2(f_3(r_{i+1},x_i))$, where  \\  $r_{i+1}=R(y_{i+1})$, $y_{i+1} =  f_1(y_i)$ } \\\hline
         $ \mF_2(x)\odot \mF_1(x) \circledcirc \mF_3(x)$ &     \tabincell{l}{ $x_{i+1} =  f_3(f_1(r_{i+1},x_i))$, where  \\  $r_{i+1}=R(y_{i+1})$, $y_{i+1} =  f_2(y_i)$ } \\\hline
         $ \mF_2(x)\odot \mF_3(x) \circledcirc \mF_1(x)$ &     \tabincell{l}{ $x_{i+1} =  f_1(f_3(r_{i+1},x_i))$, where  \\  $r_{i+1}=R(y_{i+1})$, $y_{i+1} =  f_2(y_i)$ } \\\hline
         $ \mF_3(x)\odot \mF_1(x) \circledcirc \mF_2(x)$ &     \tabincell{l}{ $x_{i+1} =  f_2(f_1(r_{i+1},x_i))$, where  \\  $r_{i+1}=R(y_{i+1})$, $y_{i+1} =  f_3(y_i)$ } \\\hline
         $ \mF_3(x)\odot \mF_2(x) \circledcirc \mF_1(x)$ &     \tabincell{l}{ $x_{i+1} =  f_1(f_2(r_{i+1},x_i))$, where  \\  $r_{i+1}=R(y_{i+1})$, $y_{i+1} =  f_3(y_i)$ } \\\hline
         $ \mF_1(x)\circledcirc \mF_2(x) \odot \mF_3(x)$ &     \tabincell{l}{ $x_{i+1} =  f_3(r_{i+1},x_i)$, where  \\  $r_{i+1}=R(y_{i+1})$, $y_{i+1} =  f_2(f_1(y_i))$ } \\\hline
         $ \mF_1(x)\circledcirc \mF_3(x) \odot \mF_2(x)$ &     \tabincell{l}{ $x_{i+1} =  f_2(r_{i+1},x_i)$, where  \\  $r_{i+1}=R(y_{i+1})$, $y_{i+1} =  f_3(f_1(y_i))$ } \\\hline
         $ \mF_2(x)\circledcirc \mF_1(x) \odot \mF_3(x)$ &     \tabincell{l}{ $x_{i+1} =  f_3(r_{i+1},x_i)$, where  \\  $r_{i+1}=R(y_{i+1})$, $y_{i+1} =  f_1(f_2(y_i))$ } \\\hline
         $ \mF_2(x)\circledcirc \mF_3(x) \odot \mF_1(x)$ &     \tabincell{l}{ $x_{i+1} =  f_1(r_{i+1},x_i)$, where  \\  $r_{i+1}=R(y_{i+1})$, $y_{i+1} =  f_3(f_2(y_i))$ } \\\hline
         $ \mF_3(x)\circledcirc \mF_1(x) \odot \mF_2(x)$ &     \tabincell{l}{ $x_{i+1} =  f_2(r_{i+1},x_i)$, where  \\  $r_{i+1}=R(y_{i+1})$, $y_{i+1} =  f_1(f_3(y_i))$ } \\\hline
         $ \mF_3(x)\circledcirc \mF_2(x) \odot \mF_1(x)$ &     \tabincell{l}{ $x_{i+1} =  f_1(r_{i+1},x_i)$, where  \\  $r_{i+1}=R(y_{i+1})$, $y_{i+1} =  f_2(f_3(y_i))$ } \\\hline
\end{tabular}
\end{center}
\end{table}

%\subsection{Discussion}
%For the combined operation defined in Eq.~\eqref{eq.co}, the operation order is from left to right. To extend the operation rules and make the generation procedure more flexible, we can import some priority symbols into the NCP model, e.g. the parenthesis. With these priority symbols, the representation can become clearer. For example, for the generation procedure
%\begin{equation*}
%\begin{split}
%  \mathcal{CO}_1 = \mF_1 \circledcirc \mF_2, \\
%  \mathcal{CO}_2 = \mF_3 \circledcirc \mF_4, \\
%  \mathcal{CO} = \mathcal{CO}_1\oplus \mathcal{CO}_2,
%\end{split}
%\end{equation*}
%we can use one equation to demonstrate it,
%\begin{equation*}
%  \mathcal{CO} = (\mF_1 \circledcirc \mF_2)\oplus (\mF_3 \circledcirc \mF_4).
%\end{equation*}
%
%Besides the six basic nonlinear operations listed in Table~\ref{tab.Operations} and their combinations, the NCP model can be further extended by defining more new nonlinear operations.

\section{Chaotic Behavior Analysis}
\label{section4}
Because chaotic behaviors are some observed phenomena, they are difficult to be qualitatively measured. Many researchers have proposed different methods to detect the existence of chaotic behavior. Among these methods, the LE developed in~\cite{Wolf1985PDNP} is a widely accepted indictor. For two close trajectories of a dynamical system, LE denotes their average divergence and it can be defined as Definition~\ref{defn.LE}.
\begin{defn}
\label{defn.LE}
  The LE of a first-order difference equation $x_{i+1}=\mF(x_i)$ is mathematically defined by
\begin{equation}\label{eq.LE}
    \lambda_{\mF(x)} = \lim_{n\rightarrow \infty}\left\{\frac{1}{n} \ln\left|\frac{\mF^n(x_0+\epsilon)-\mF^n(x_0))}{\epsilon}\right|\right\},
\end{equation}
where $\epsilon$ is a very small positive value. If $\mF(x)$ is differentiable, Eq.~\eqref{eq.LE} can be rewritten as
\begin{equation}\label{eq.LER}
   \lambda_{\mF(x)}  = \lim_{n\rightarrow \infty}\left\{\frac{1}{n} \sum_{i=0}^{n-1} \ln\left|\mF'(x_i)\right|\right\}.
\end{equation}
\end{defn}

A positive LE denotes that the two close trajectories of a dynamical system exponentially diverge in each unit time and they will be totally different eventually, while a negative LE means that their distance reduces and they will finally overlap as time goes to infinity. Thus, a dynamical system $x_{i+1}=\mF(x_i)$ is considered to own chaotic behavior if $\lambda_{\mF(x)}>0$.

Using the LE theory, this section analyzes the chaotic behaviors of the NCP model. The chaotic behaviors of the cascade, modulation, switching and fusion operations have been analyzed in our previous work in~\cite{Zhou2015IETC,Hua2016IETC,Wu2014IETCS1,Zhou2014SP}. Theoretical analysis and experimental results showed that all the cascade, modulation, switching and fusion operations can generate chaotic maps with complex chaotic behaviors.
\subsection{Chaotic Behavior of Scalar Cascade}
For the scalar cascade operation $\mathcal{U}(x)=c\ \widetilde{\circledcirc}\ \mF(x)$, when $c=2$, its iterative form can be rewritten as $x_{i+1}=\mF(\mF(x_i))$. Based on the definition of LE in Eq.~\eqref{eq.LER}, its LE can be written as
\begin{equation*}
\begin{split}
   \lambda_{\mathcal{U}(x)} = & \lim_{n\rightarrow \infty}\left\{\frac{1}{n} \sum_{i=0}^{n-1} \ln\left|(\mF(\mF(x_i)))'\right|\right\} \\
    = & \lim_{n\rightarrow \infty}\left\{\frac{1}{n} \sum_{i=0}^{n-1} \ln\left| \mF'(\mF(x_i))\mF'(x_i)\right| \right\} \\
    = &  \lim_{n\rightarrow \infty}\left\{\frac{1}{n}\sum_{i=0}^{n-1}\ln\left|\mF'(\mF(x_i))\right|\right\}  \\
     & + \lim_{n\rightarrow \infty}\left\{\frac{1}{n}\sum_{i=0}^{n-1}\ln\left|\mF'(x_i)\right|\right\} \\
    = & 2\lambda_{\mF(x)}.
\end{split}
\label{FxLYP}
\end{equation*}
When $c=k$, it is not difficult to calculate out that $\lambda_{\mathcal{U}(x)} = k\lambda_{\mF(x)}$. Thus, $\lambda_{\mathcal{U}(x)} >0$ if $\lambda_{\mF(x)}>0$. This means that if the seed map $\mF(x)$ has chaotic behavior, the scalar cascade result is chaotic and has larger LE than its seed map.

\subsection{Chaotic Behavior of Scalar Modulation}
\label{section.scamoCha}
Based on the definition of LE in Eq.~\eqref{eq.LER}, the LE of the scalar modulation shown in Fig.~\ref{fig.ScMo} can be written as
\begin{equation}\label{eq.LESM1}
   \lambda_{\mathcal{D}(x)}  = \lim_{n\rightarrow \infty}\left\{\frac{1}{n} \sum_{i=0}^{n-1} \ln\left|\mF'(r_{i+1},x_i)\right|\right\},
\end{equation}
where $x_i$ is the iteration value and $r_{i+1}$ is the transformation result of $(c-1)$-th control map $\mF(y^{(c-1)})$'s output that is used to control the parameter of $\mF(x)$ in each iteration. Because the parameter of $\mF(y^{(c-1)})$ is modulated by $\mF(y^{(c-2)})$, $\mF(y^{(c-1)})$ has chaotic attractor based on the analysis in Section
\uppercase\expandafter{\romannumeral3}-C of~\cite{Hua2016IETC}, Then, the seed map $\mF(x)$ achieves a different control parameter in each iteration to make the iterative outputs different and unpredictable.

\subsection{Chaotic Behaviors of Combination Operations}
To demonstrate the chaotic behavior of the combination in Eq.~\eqref{eq.co}, we analyze the chaotic behaviors of the combination operations in Table~\ref{tab.Eg}. According to the order of the two basic operations, the 12 combination operations in Table~\ref{tab.Eg} can be classified into two kinds. One is first do the modulation and then performs the cascade. The other is first do the cascade and then performs the modulation. For the two kinds of operations, we separately take one example to analyze its chaotic behavior. The two examples are $\mathcal{CO}_1(x)=\mF_1(x)\odot \mF_2(x) \circledcirc \mF_3(x)$ and $\mathcal{CO}_2(x)=\mF_1(x)\circledcirc  \mF_2(x) \odot \mF_3(x)$.

\subsubsection{Chaotic Behavior of $\mathcal{CO}_1(x)$}
The example $\mathcal{CO}_1(x)$ first performs the modulation to $f_1(x)$ and $f_2(x)$, and then cascades the modulation result and $f_3(x)$. Suppose $\mathcal{M}(x)=f_1(x)\odot f_2(x)$. According to the analysis in Section
\uppercase\expandafter{\romannumeral3}-C of~\cite{Hua2016IETC} that if the seed map $f_2(x)$ has continuous chaotic range, the modulation result $\mathcal{M}(x)$ always has chaotic behavior. This means that $\lambda_{\mathcal{M}(x)}>0$.

The example $\mathcal{CO}_1(x)$ can be rewritten as $\mathcal{CO}_1(x)=\mathcal{M}(x)\circledcirc \mF_3(x)$, namely $x_{i+1}=f_3(\mathcal{M}(x_i))$. Based on the definition of LE in Eq.~\eqref{eq.LER}, the LE of $\mathcal{CO}_1(x)$ can be defined as

\begin{equation*}
\begin{split}
   \lambda_{\mathcal{CO}_1(x)} = & \lim_{n\rightarrow \infty}\left\{\frac{1}{n} \sum_{i=0}^{n-1} \ln\left|(f_3(\mathcal{M}(x_i)))'\right|\right\} \\
    = & \lim_{n\rightarrow \infty}\left\{\frac{1}{n} \sum_{i=0}^{n-1} \ln\left| f_3'(\mathcal{M}(x_i))\mathcal{M}'(x_i)\right| \right\} \\
    = &  \lim_{n\rightarrow \infty}\left\{\frac{1}{n}\sum_{i=0}^{n-1}\ln\left|f_3'(\mathcal{M}(x_i))\right|\right\}  \\
     & + \lim_{n\rightarrow \infty}\left\{\frac{1}{n}\sum_{i=0}^{n-1}\ln\left|\mathcal{M}'(x_i)\right|\right\} \\
    = & \lambda_{\mF_3(x)}+\lambda_{\mathcal{M}(x)}.
\end{split}
\label{FxLYP}
\end{equation*}
Thus, the LE of $\mathcal{CO}_1(x)$ is the combination of those of $\mathcal{M}(x)$ and $f_3(x)$. Because $\lambda_{\mathcal{M}(x)}>0$, $\lambda_{\mathcal{CO}_1(x)}>0$. This means that $\mathcal{CO}_1(x)$ is chaotic.

\subsubsection{Chaotic Behavior of $\mathcal{CO}_2(x)$}
The example $\mathcal{CO}_2(x)$ first cascades $f_1(x)$ and $f_2(x)$, and then performs modulation to the cascade result and $f_3(x)$.
Suppose $\mathcal{C}(x)=\mF_1(x)\circledcirc \mF_2(x)$, then $\mathcal{CO}_2(x)=\mathcal{C}(x) \odot f_3(x)$.
Based on the definition of LE in Eq.~\eqref{eq.LER}, the LE of $\mathcal{CO}_2(x)$ can be written as
\begin{equation}\label{eq.LESM1}
   \lambda_{\mathcal{CO}_2(x)}  = \lim_{n\rightarrow \infty}\left\{\frac{1}{n} \sum_{i=0}^{n-1} \ln\left|\mF_3'(r_{i+1},x_i)\right|\right\},
\end{equation}
where $x_i$ is the iteration value and $r_{i+1}$ is the transformation result of $\mathcal{C}(x)$'s output that is used to control the parameter of $\mF_3(x)$ in each iteration. Its LE can be analyzed from the following ways:
\begin{itemize}
  \item When the attractors of $\mathcal{C}(x)$ are an equilibrium point, after transforming, the obtained $r_{i+1}$ is also fixed and within the chaotic range of $\mF_3(x)$. Thus, $$\lambda_{\mathcal{CO}_2(x)} > 0.$$
  \item When the attractors of $\mathcal{C}(x)$ are a limit cycle, namely, $\mathcal{C}(x)$ has a periodic orbit and its outputs are a finite number of different points, suppose $\{o_j~|~j=1,2,\cdots,k\}$. After transforming, the periodic sequence $\{o_j~|~j=1,2,\cdots,k\}$ is transformed as $\{r_j~|~j=1,2,\cdots,k\}$, which is also a periodic sequence and $r_j$~$(j=1,2,\cdots,k)$ is in the chaotic range of $\mF_3(x)$. Because $k$ is a finite number, when the iteration number $n$ increases to $\infty$, the number of each point of the periodic sequence $\{r_j~|~j=1,2,\cdots,k\}$ approaches to $n/k$. Thus, Eq.~\eqref{eq.LESM1} can be rewritten as
      \begin{equation}
      \label{eq.LESM2}
      \begin{split}
        \lambda_{\mathcal{CO}_2(x)} = & \lim_{n\rightarrow \infty}\left\{\frac{1}{n} \sum_{i=0}^{n/k-1} \ln\left|(\mF_3'(r_1,x_i)\right|\right\}+\cdots   \\
         &+\lim_{n\rightarrow \infty}\left\{\frac{1}{n} \sum_{i=0}^{n/k-1} \ln\left|(\mF_3'(r_k,x_i)\right|\right\}.
        \end{split}
      \end{equation}

      Because $k$ is a finite number and $n\rightarrow\infty$, then $(n/k)\rightarrow\infty$. Thus,
      \begin{equation*}
      \begin{split}
         &\lim_{n\rightarrow \infty}\left\{\frac{1}{n}\sum_{i=0}^{n/k-1} \ln\left|(\mF_3'(r_j,x_i)\right|\right\} \\
         &= \frac{1}{k}\lim_{(n/k)\rightarrow \infty}\left\{\frac{1}{n/k} \sum_{i=0}^{n/k-1} \ln\left|(\mF_3'(r_j,x_i)\right|\right\} \\
         &= \frac{1}{k}\lambda_{\mF_3(r_j,x)},
        \end{split}
      \end{equation*}
      where $j=1,2,\cdots,k$. Then, Eq.~\eqref{eq.LESM2} becomes
       \begin{equation*}\label{kparameters}
        \begin{split}
          \lambda_{\mathcal{CO}_2(x)}  = &\frac{1}{k}\lambda_{\mF_3(r_1,x)} + \frac{1}{k}\lambda_{\mF_3(r_2,x)}+\cdots+\frac{1}{k}\lambda_{\mF_3(r_k,x)} \\
           = &\frac{1}{k}\sum_{j=1}^{k}\lambda_{\mF_3(r_j,x)}.
        \end{split}
        \end{equation*}

      Because $r_j~(j=1,2,\cdots,k)$ is in the chaotic range of $\mF_3(x)$, $\lambda_{\mF_3(r_j,x)}>0$ for $~\forall j\in\{1,2,...,k\}$. Thus, $$\lambda_{\mathcal{CO}_2(x)} = \frac{1}{k}\sum_{j=1}^{k}\lambda_{\mF_3(r_j,x)} > 0.$$

  \item When $\mathcal{C}(x)$ has chaotic attractor, $\mathcal{C}(x)$ is chaotic and $r_{i+1}$ is dynamical. In this case, the seed map $\mF_3(x)$ achieves a different control parameter in each iteration to make the iterative outputs different and unpredictable.
\end{itemize}
In summary, $\mathcal{CO}_2(x)$ always has chaotic behavior.

\section{Examples of New Chaotic Maps}
\label{Section5}
This section demonstrates four examples of chaotic maps generated by the NCP model.

\subsection{$\mE_1$}
First, we demonstrate the scalar cascade operation $\widetilde{\circledcirc}$. The constant $c$ is set as 3 and the seed map is selected as the Sine map $\mS(x)$, then a chaotic map $\mE_1$ can be generated. Mathematically, it is represented as
\begin{equation*}
  \begin{split}
    x_{i+1} & = 3 \ \widetilde{\circledcirc}\ \mS(x_i) \\
            & = p_1\sin(\pi p_2\sin(\pi p\sin(\pi x_i))),
  \end{split}
\end{equation*}
where $p,~p_1,~p_2$ are control parameters within the range $[0,1]$. For simplify, we set the parameters $p_1=1$, $p_2=1$ and investigate the chaotic behavior of $\mE_1$ with the change of its parameter $p$. Then,
\begin{equation}
  x_{i+1}=\sin(\pi \sin(\pi p\sin(\pi x_i))).
\end{equation}

\subsubsection{Equilibrium point and stability}
To find out the equilibrium points of $\mE_1$, we set $x_{i+1}=x_i$ and the equilibrium points of $\mE_1$ are the roots of the equation
\begin{equation}\label{eq.fE1}
   \tilde{x}-\sin(\pi \sin(\pi p\sin(\pi \tilde{x}))) = 0.
\end{equation}
Obviously, $\tilde{x}_1=0$ is one equilibrium point of $\mE_1$. Solving Eq.~\eqref{eq.fE1}, we can find out that $\mE_1$ has more equilibrium points when its control parameter $p$ increases within the range $[0,1]$. When $p>0.0323$, Eq.~\eqref{eq.fE1} has another root $\tilde{x}_2$ and thus $\mE_1$ has two equilibrium points; When $p>0.3063$, $\mE_1$ has four equilibrium points; When $p>0.6831$, $\mE_1$ has six equilibrium points; When $p>0.9460$, the number of equilibrium points increases to eight.

The equilibrium point of a dynamical system has two states: stable and unstable. Its stability is dependent on the slope of the system' curve at the point. When the slope is within the range $(-45^{\circ},45^{\circ})$, the equilibrium point is stable and it attracts all its neighboring trajectories to make them converge to the point eventually; otherwise, the equilibrium point is unstable and its neighboring trajectories escape from it as the time increases. The Jacobian matrix can be used to calculate the slope of a curve and that of $\mE_1$ is given by
\begin{equation*}\label{eq.E1}
\begin{split}
  J = &\ \frac{d_{\mE_1}}{d_{x_i}}  \\
    = &\ \cos(\pi \sin(\pi p\sin(\pi x_i))) \\
      &\ \pi \cos(\pi p\sin(\pi x_i))\pi p \cos(\pi x_i)\pi.
  \end{split}
\end{equation*}
When the Jacobian value at the point is within the range $(-1,1)$, the corresponding slope falls into the range $(-45^{\circ},45^{\circ})$ to make the equilibrium point stable; otherwise, the equilibrium point is unstable.

\begin{figure}[htbp]
\centering
  \begin{minipage}[b]{0.98\linewidth}
    \begin{minipage}[b]{0.49\linewidth}
      \centerline{\includegraphics[width=1\linewidth]{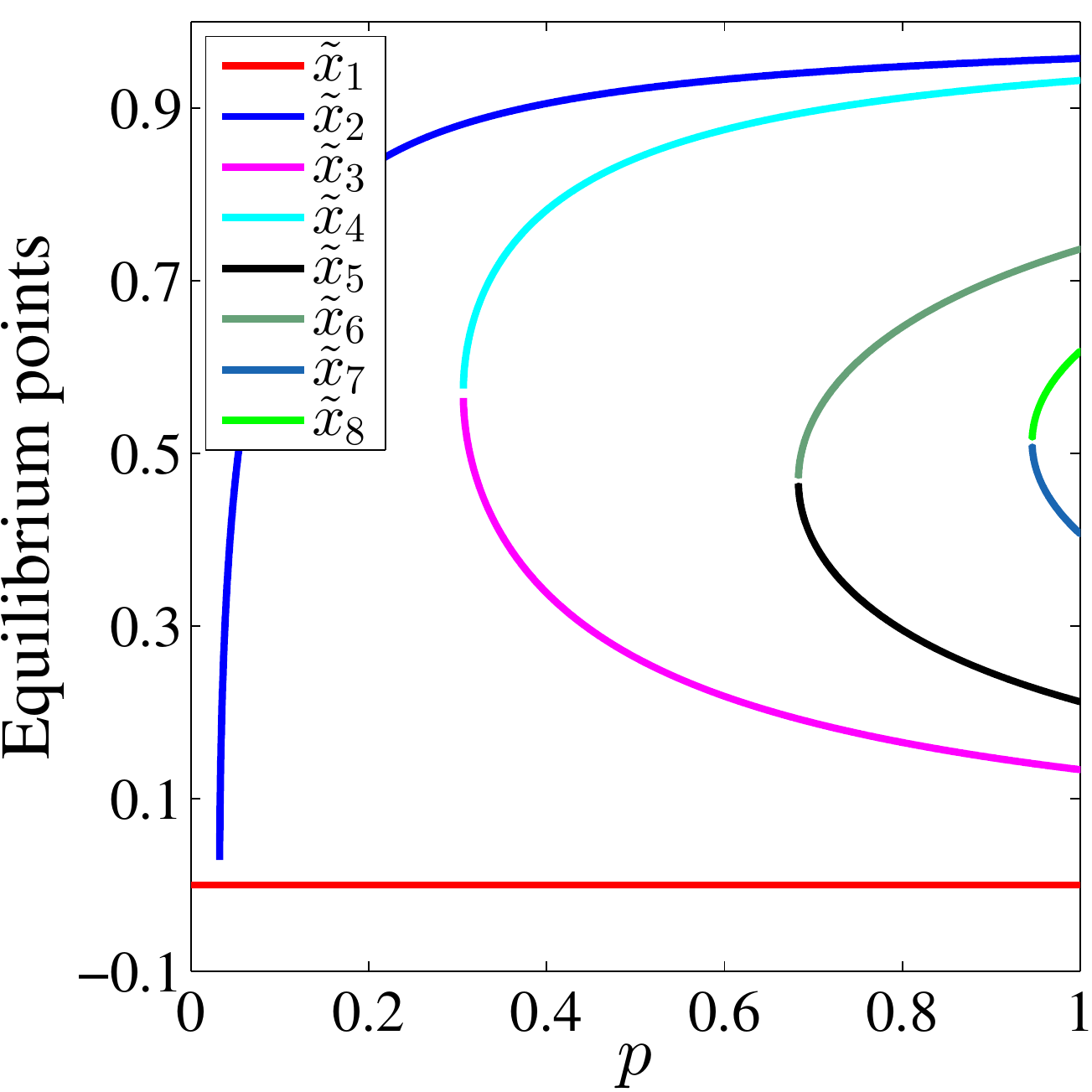}}
      \centerline{(a)}
    \end{minipage}\hfill
      \begin{minipage}[b]{0.49\linewidth}
       \centerline{\includegraphics[width=1\linewidth]{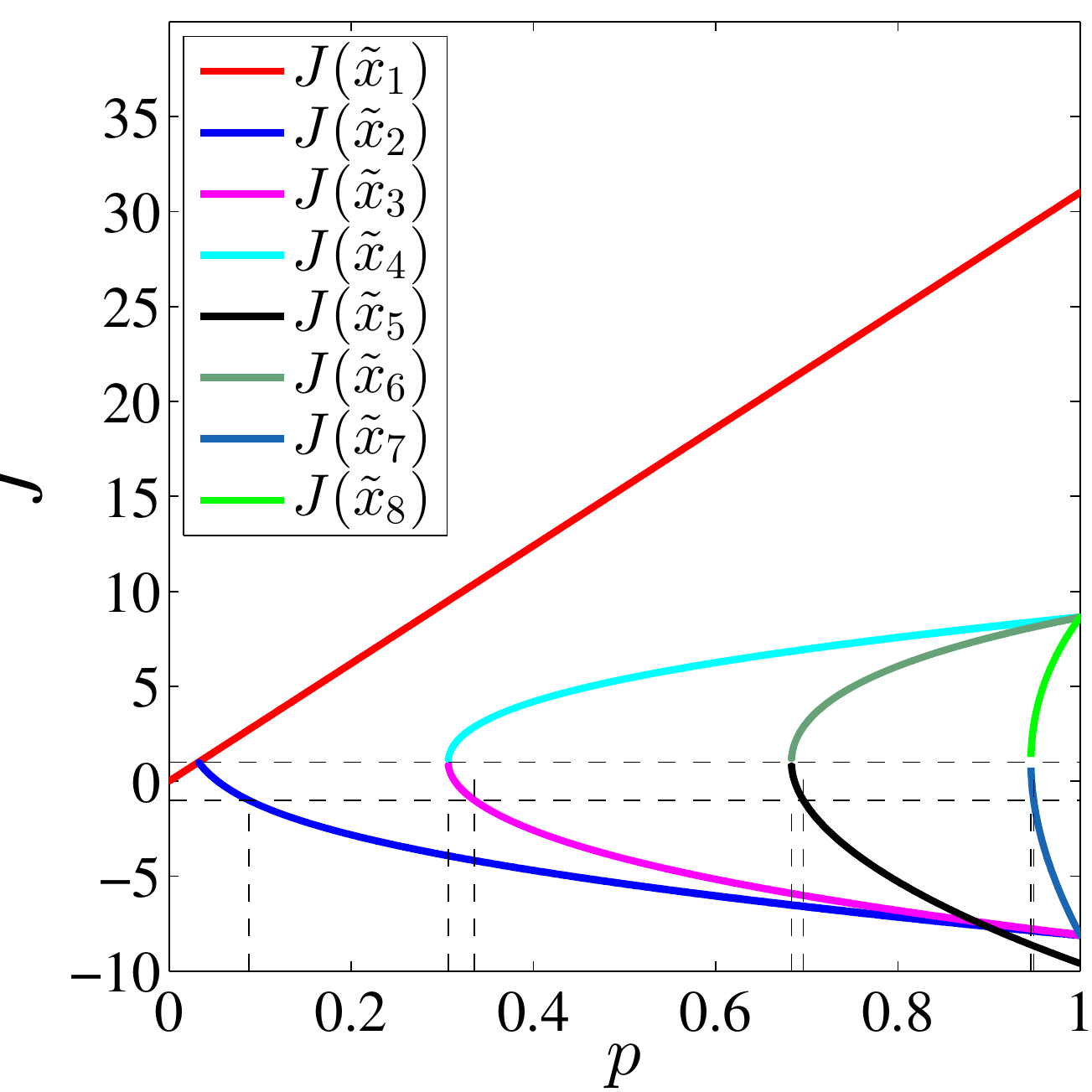}}
       \centerline{(b)}
      \end{minipage}\hfill
  \end{minipage}\hfill
    \caption{(a) Equilibrium points of $\mE_1$; (b) Jacobian values of the equilibrium points of $\mE_1$.}\label{fig.E1}
\end{figure}

Fig.~\ref{fig.E1} plots the equilibrium points of $\mE_1$ and their Jacobian values with different parameter settings. Table~\ref{tab.Fix2} lists the occurrence and stable intervals of these equilibrium points. As can be seen from the table, we can achieve that $\mE_1$ has stability when $p\in[0,0.0875]\cup[0.3063,0.3352]\cup[0.6831,0.6961]\cup[0.9460,0.9490]$.

\begin{table}[htbp]
\newcommand{\tabincell}[2]{\begin{tabular}{@{}#1@{}}#2\end{tabular}}
\renewcommand{\arraystretch}{1.5}
\setlength{\tabcolsep}{3pt}
\begin{center}
\caption{Equilibrium points of $\mE_1$ and their stability.}\label{tab.Fix2}
\begin{tabular}{|c|c|c|} \hline
 Equilibrium points & Occurrence intervals & Stable intervals \\\hline
 $\tilde{x}_1$ & $[0,1]$ & $[0,0.0323]$ \\\hline
  $\tilde{x}_2$ & $[0.0323,1]$ & $[0.0323,0.0875]$ \\\hline
   $\tilde{x}_3$ & $[0.3063,1]$ & $[0.3063,0.3352]$ \\\hline
    $\tilde{x}_4$ & $[0.3063,1]$ & unstable \\\hline
     $\tilde{x}_5$ & $[0.6831,1]$ & $[0.6831,0.6961]$ \\\hline
      $\tilde{x}_6$ & $[0.6831,1]$ & unstable\\\hline
       $\tilde{x}_7$ & $[0.9460,1]$ & $[0.9460,0.9490]$ \\\hline
        $\tilde{x}_8$ & $[0.9460,1]$ & unstable \\\hline

\end{tabular}
\end{center}
\end{table}

\subsubsection{Bifurcation diagram}
\begin{figure}[htbp]
\centering
  \begin{minipage}[b]{0.52\linewidth}
    \centerline{\includegraphics[width=1\linewidth]{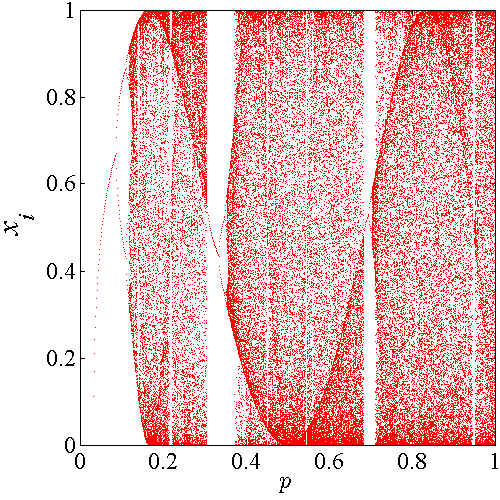}}
  \end{minipage}\hfill
    \caption{The bifurcation diagram of $\mE_1$.}
    \label{fig.E1Bifur}
\end{figure}

As can be observed from Fig.~\ref{fig.E1}(b), when the control parameter $p$ increases to $0.0875$, the equilibrium point $\tilde{x}_2=0.6708$ becomes unstable. At the same time, the Jacobian value at this point reduces to -1, which means that the slope of the tangent line of the system steepening to $-45^\circ$. Once this happens, the equilibrium point $\tilde{x}_2$ becomes two new stable points. When $p$ increases to $0.1109$, the period-two stable points lose their stability and generate period-four stable points; When $p$ increases to $0.1172$, the period-four stable points lose their stability and generate eight stable points. For example, when $p=0.1185$, the eight stable points are $0.3521$, $0.8554$, $0.4890$, $0.9095$, $0.3217$, $0.8274$, $0.5646$, $0.9002$. By this principle, the stable points doubly increase and a critical value $\tilde{p}$ is finally achieved. When $p$ is slightly less than $\tilde{p}$, the outputs of the system are periodic with a large period. When $p$ is slightly larger than $\tilde{p}$, these points start to become aperiodic and the system eventually route to chaos, which is called period-doubling bifurcation. Numerical result shows that $\hat{p}=0.1190$. When $p$ increases to $0.3063$, $0.6831$ or $0.9460$, $\mE_1$ obtains the stable equilibrium points $\tilde{x}_3$, $\tilde{x}_5$, $\tilde{x}_7$, respectively. Then, it returns back to stable state. When $p$ increases to $0.3352$, $0.6961$ or $0.9490$, $\tilde{x}_3$, $\tilde{x}_5$ and $\tilde{x}_7$ loss their stability and $\mE_1$ starts to route to chaos again. The bifurcation diagram of $\mE_1$ is plotted in Fig.~\ref{fig.E1Bifur}.

% which indicates that the system has chaotic behaviors when $p\in[0.1190,0.3063]\cup[0.3543,0.6831]\cup[0.7043,0.9460]\cup[0.9509,1]$.

\subsection{$\mE_2$}
Here, we give an example of chaotic map generated by the scalar modulation operation $\widetilde{\odot}$. The coefficient $c$ is set as 3 and the seed map is also selected as the Sine map $\mS(x)$, then a chaotic map $\mE_2$ can be generated by
\begin{equation}
  \begin{split}
  x_{i+1} & = 3\ \widetilde{\odot}\ \mS(x_i) \\
          & =  r_{i+1}\sin(\pi x_i),
  \end{split}
\end{equation}
where
\begin{equation}
  \begin{split}
   r_{i+1} &= 1-0.13y_{i+1}^{(2)}, \\
   y_{i+1}^{(2)} &= {p}_{i+1}\sin(\pi y_{i}^{(2)}),\\
   p_{i+1} &= 1-0.13{y}_{i+1}^{(1)}, \\
   {y}_{i+1}^{(1)} &= p\sin(\pi {y}_{i}^{(1)}),
\end{split}
\end{equation}
where $p$ is the control parameter and $p\in[0,1]$.

\begin{figure}[htbp]
\centering
  \begin{minipage}[b]{0.52\linewidth}
    \centerline{\includegraphics[width=1\linewidth]{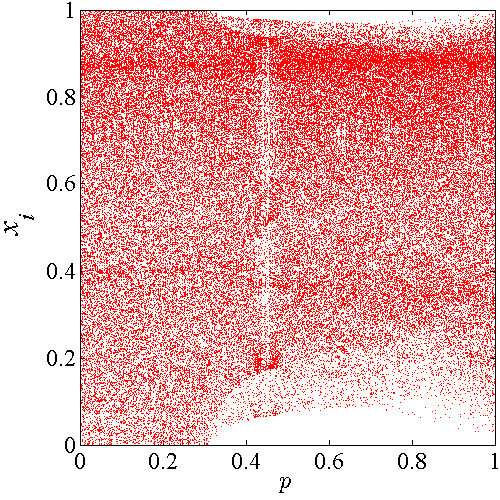}}
  \end{minipage}\hfill
    \caption{The bifurcation diagram of $\mE_2$.}
    \label{fig.E2Bifur}
\end{figure}

Fig.~\ref{fig.E2Bifur} shows the bifurcation diagram of $\mE_2$. Theoretically, if the seed map $\mF(x)$ has a continuous chaotic range, the scalar modulation result has chaotic behavior for all the parameter settings. This has been proved in Section~\ref{section.scamoCha}. However, if the chaotic range of $\mF(x)$ is not continuous, the scalar modulation result may loss its chaotic behavior in some parameter settings. This occurs when the fixed outputs of a control map happen to be transformed into the non-chaotic ranges of the next control map or seed map, such as the white space in the chaotic ranges of the Logistic and Sine maps (see Figs.~\ref{Fig.ExistingMaps}(a) and (b)). As the seed map in $\mE_2$ is the Sine map, $\mE_2$ losses its chaotic behavior in few parameter settings, which can be observed from Fig.~\ref{fig.E2Bifur}.

\subsection{$\mE_3$}
The combination operations in the NCP model can also generate chaotic maps with complex chaotic behavior. This example demonstrates the operation $\mathcal{CO}_1(x)=\mF_1(x)\odot \mF_2(x) \circledcirc \mF_3(x)$ in Table~\ref{tab.Eg}. $f_1(x)$ is selected as the Tent map $\mT(x)$; $f_2(x)$ is selected as the Sine map $\mS(x)$; and $f_3(x)$ is selected as the Logistic map $\mL(x)$. Then, $\mE_3$ is defined by
\begin{equation*}
  \mE_3(x)= \mT(x) \odot \mS(x) \circledcirc \mL(x).
\end{equation*}
First, Tent map controls the parameter of the Sine map to generate a new chaotic map $\mathcal{M}(x)$. Then, $\mathcal{M}(x)$ is cascaded by the Logistic map to obtain $\mE_3$. The iterative form of $\mE_3$ is represented as
\begin{equation}
  x_{i+1} = 4r_{i+1}\sin(\pi x_i)(1-r_{i+1}\sin(\pi x_i)),
\end{equation}
where $r_{i+1}$ is the transformation result of $y_{i+1}$, which is defined as
\begin{equation}
r_{i+1} = 1-0.13y_{i+1},
\end{equation}
where $y_{i+1}$ is the output of the Tent map,
\begin{equation*}
  y_{i+1}=\begin{cases} 2py_1, &\mbox{for}\ y_i < 0.5, \\
  2p(1-y_i),  &\mbox{for}\ y_i\geq 0.5, \end{cases}
\end{equation*}
where the control parameter $p\in[0,1]$.

\begin{figure}[htbp]
\centering
  \begin{minipage}[b]{0.52\linewidth}
    \centerline{\includegraphics[width=1\linewidth]{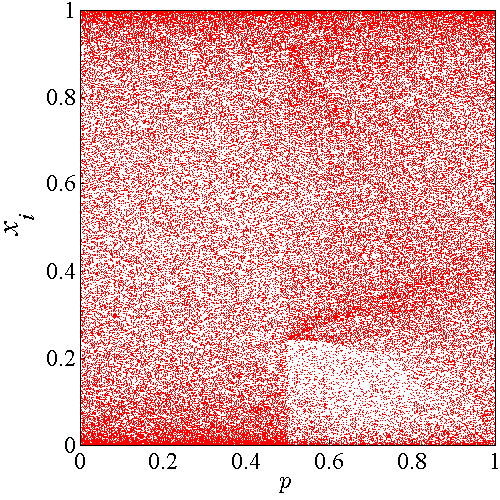}}
  \end{minipage}\hfill
    \caption{The bifurcation diagram of $\mE_3$.}
    \label{fig.E3Bifur}
\end{figure}

We use the numerical result to investigate the chaotic properties of $\mE_3$. It is obvious that $\tilde{x}_1=0$ is an equilibrium point of $\mE_3$. In the generation procedure of $\mE_3$, the parameter of Sine map is modulated by the Tent map. When $p\in[0,0.5]$, the Tent map has fixed point, which is transformed into the chaotic range of Tent map by the following transformation; When $p\in(0.5,1]$, the Tent map has chaotic attractors and the Sine map gets a dynamically changed parameter in each iteration. After cascading with the Logistic map, the obtained $\mE_3$ is dissipated in the whole parameter range $[0,1]$, which is also verified by its bifurcation diagram shown in Fig.~\ref{fig.E3Bifur}.

\subsection{$\mE_4$}
The example $\mE_4$ demonstrates the operation $\mathcal{CO}_2(x)=\mF_1(x) \circledcirc \mF_2(x) \odot \mF_3(x)$ in Table~\ref{tab.Eg}. As any different of same chaotic maps can be selected in the combination operation, we select $f_1(x)$ and $f_2(x)$ both as the Tent map $\mT(x)$ and choose $f_3(x)$ as the Sine map $\mS(x)$. Then, $\mE_4$ is defined by
\begin{equation*}
\begin{split}
  \mE_4(x) = \mT(x) \circledcirc \mT(x) \odot \mS(x).
  \end{split}
\end{equation*}
First, Tent map is cascaded to itself to generate a new chaotic map, namely $\mathcal{C}(x)=\mT(x) \circledcirc \mT(x)$. Then, $\mathcal{C}(x)$ is to dynamically control the parameter of the Sine map to obtain $\mE_4$. The iterative definition of $\mE_4$ can be represented by
\begin{equation}
  x_{i+1} = r_{i+1}\sin(\pi x_i),
\end{equation}
where $r_{i+1}$ is the transformation result of $y_{i+1}$, which is defined as
\begin{equation}
r_{i+1} = 1-0.13y_{i+1},
\end{equation}
where $y_{i+1}$ is the output of $\mathcal{C}(x)$.

\begin{figure}[htbp]
\centering
  \begin{minipage}[b]{0.52\linewidth}
    \centerline{\includegraphics[width=1\linewidth]{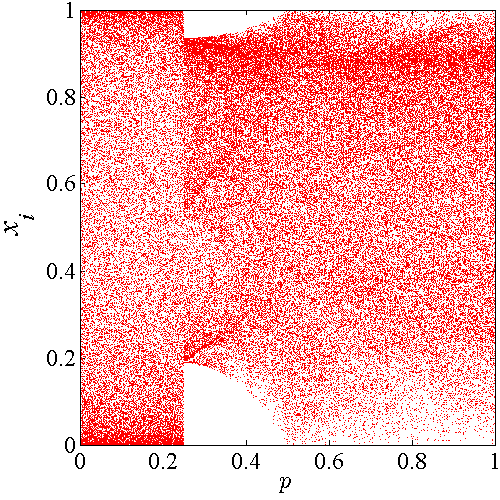}}
  \end{minipage}\hfill
    \caption{The bifurcation diagram of $\mE_4$.}
    \label{fig.E4Bifur}
\end{figure}

The bifurcation diagram of $\mE_4$ is shown in Fig.~\ref{fig.E4Bifur}, from which we can observe that $\mE_4$ also has chaotic behavior in the whole parameter range.

\section{Performance Analysis}
\label{section6}
This section evaluates the performance of the four new chaotic maps from four aspects: LE, SE~\cite{Lin1991IETIT}, CD~\cite{Grassberger1983PRL} and initial state sensitivity.
\subsection{Lyapunov Exponent}
\begin{figure}[htbp]
\centering
\begin{minipage}[b]{0.98\linewidth}
  \begin{minipage}[b]{1\linewidth}
    \centerline{\includegraphics[width=1\linewidth]{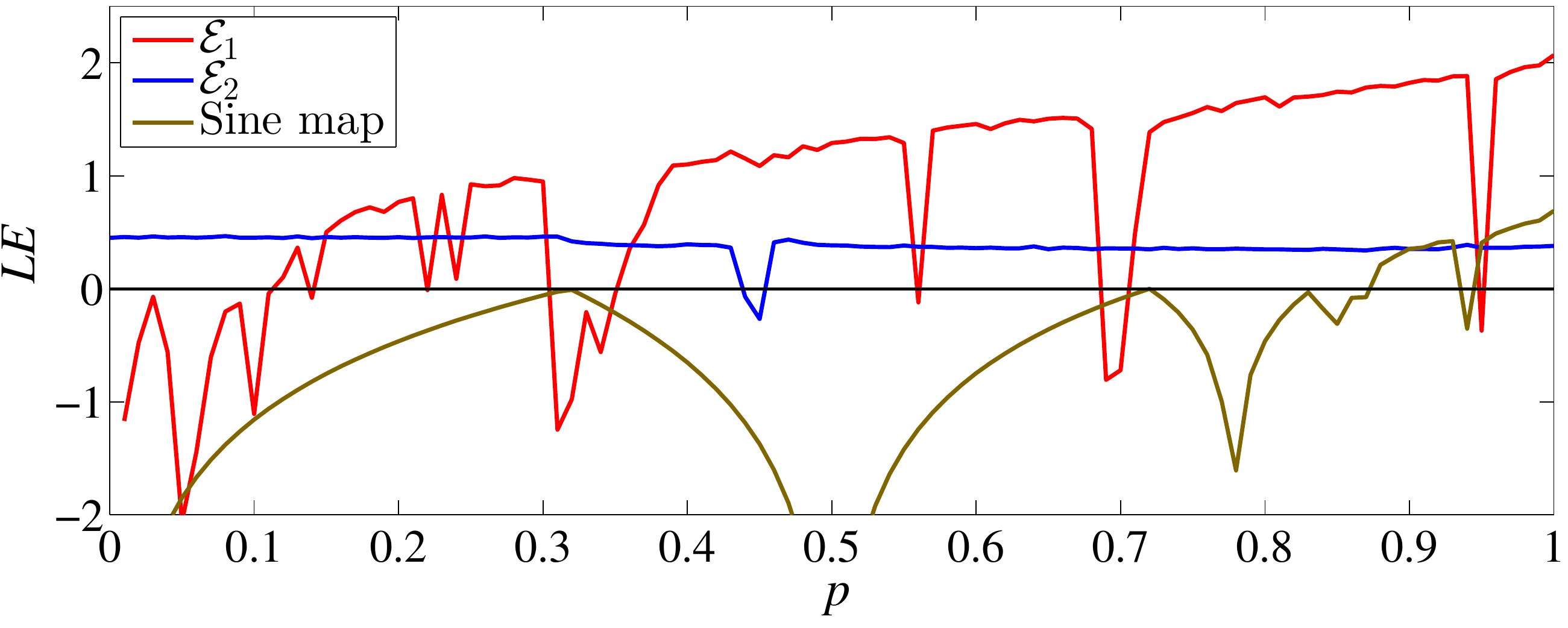}}
    \centerline{(a)}
  \end{minipage}\hfill
  \begin{minipage}[b]{1\linewidth}
    \centerline{\includegraphics[width=1\linewidth]{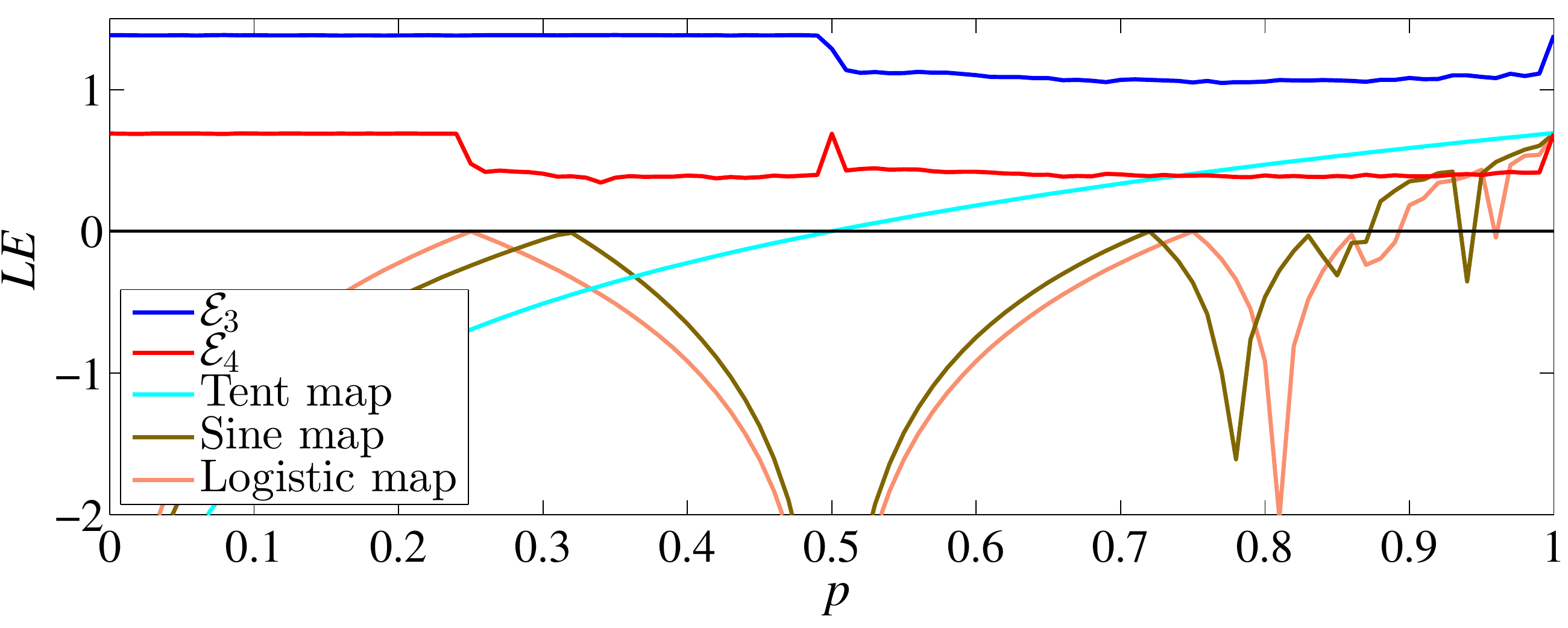}}
    \centerline{(b)}
    \end{minipage}\hfill
\end{minipage}\hfill
    \caption{(a) LE comparisons of $\mE_1$, $\mE_2$ and Sine map; (b) LE comparisons of $\mE_4$, $\mE_3$, Tent, Sine and Logistic maps.}
    \label{fig.LEs}
\end{figure}

As discussed in Section~\ref{section4} that LE is a widely accepted indictor to measure the existence of chaotic behavior. A dynamical system with at least one positive LE shows complicated dynamics and bigger positive LE means that the two close trajectories of a dynamical system diverge faster. Fig.~\ref{fig.LEs} plots the LEs of different chaotic maps with the change of their parameters. As can be observed from the figure, $\mE_1$ and $\mE_2$ have positive LEs in most parameter settings while $\mE_3$ and $\mE_4$ have positive LEs in the whole parameter ranges. This is consistent with their bifurcation diagrams in Figs.~\ref{fig.E1Bifur},~\ref{fig.E2Bifur},~\ref{fig.E3Bifur} and~\ref{fig.E4Bifur}. Compared with their corresponding seed maps used in the generation procedures, $\mE_1$, $\mE_2$, $\mE_3$ and $\mE_4$ have bigger positive LEs in most parameter settings. This means that the NCP model can generate chaotic maps with more complicated behaviors.

\subsection{Shannon Entropy}
The SE is a widely used standard to measure the randomness of a data sequence or a signal, which is defined by Definition~\ref{defn.SE}.
\begin{defn}
\label{defn.SE}
  The SE of a data sequence or a time series $\textbf{z}$ is defined as
  \begin{equation}
  H(\textbf{z}) = -\sum_{i=1}^{L}Pr(i)\log_2Pr(i),
  \label{eq.SH}
\end{equation}
where $L$ denotes the number of possible values and $Pr(i)$ represents the probability of the $i$-th possible value in $\textbf{z}$.
\end{defn}
A larger SE indicates that the values in $\textbf{z}$ distribute more random, and $H_{\max}=\log_2L$ if and only if $Pr(i)=1/L$ for $\forall i\in[1,L]$.

\begin{figure}[htbp]
\centering
\begin{minipage}[b]{0.98\linewidth}
  \begin{minipage}[b]{1\linewidth}
    \centerline{\includegraphics[width=1\linewidth]{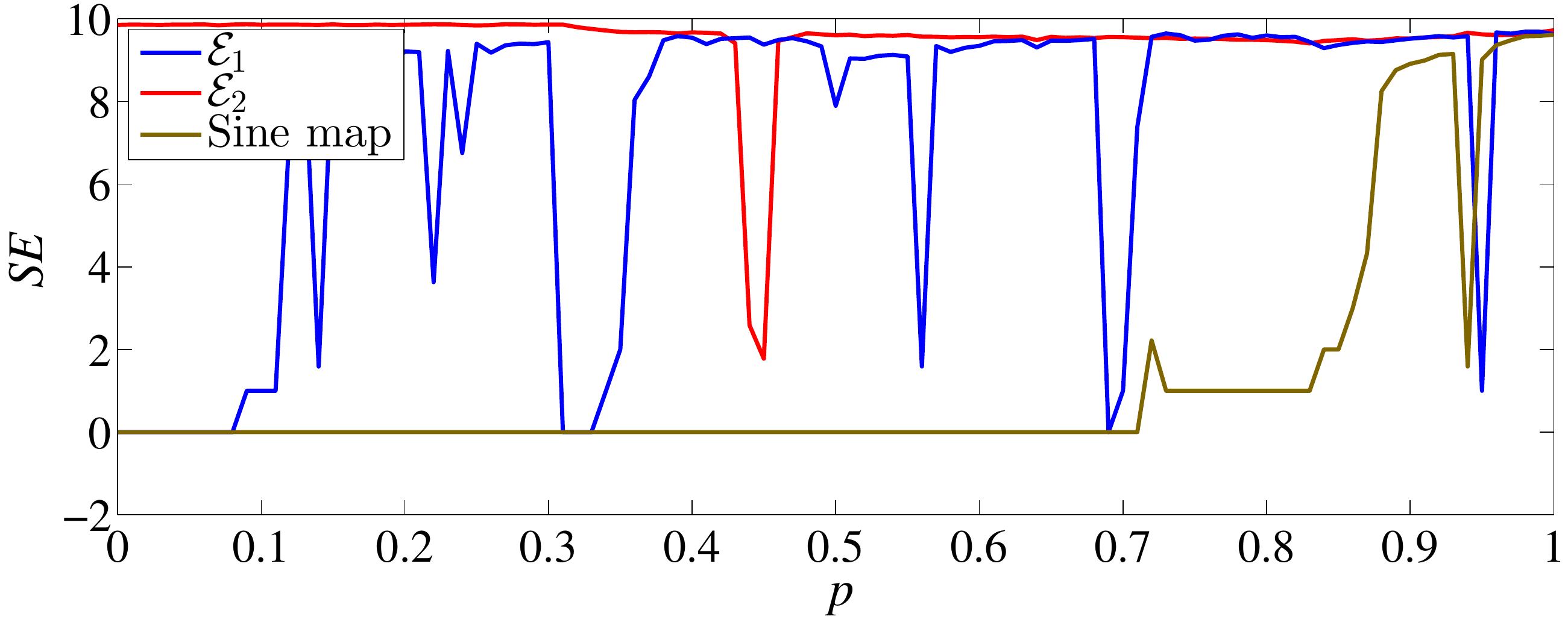}}
    \centerline{(a)}
  \end{minipage}\hfill
  \begin{minipage}[b]{1\linewidth}
    \centerline{\includegraphics[width=1\linewidth]{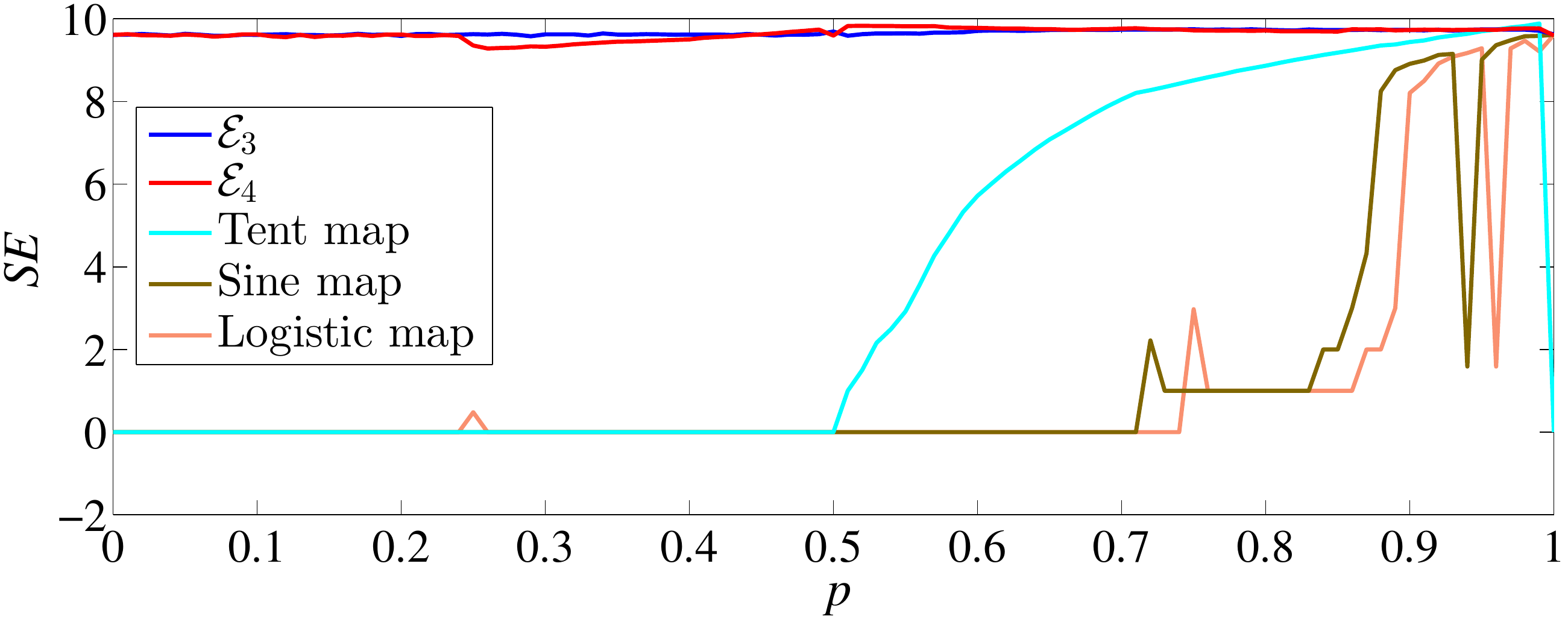}}
    \centerline{(b)}
    \end{minipage}\hfill
\end{minipage}\hfill
    \caption{(a) SE comparisons of $\mE_1$, $\mE_2$ and Sine map; (b) SE comparisons of $\mE_4$, $\mE_3$, Tent, Sine and Logistic maps.}
    \label{fig.SHs}
\end{figure}

To test the randomness of outputs of different chaotic maps, we designed the following experiments for each chaotic map: 1) obtain a time series $\textbf{z}$ with length 10,000 for different parameter settings; 2) uniformly divide interval $(0,1)$ into $2^{10}$ sub-intervals and $Pr(i)$ is the frequency of occurrence of $\textbf{z}$ in the $i$-th sub-interval; 3) calculate SE of $\textbf{z}$ using Eq.~\eqref{eq.SH}. Fig.~\ref{fig.SHs} plots SEs of different chaotic maps with different parameter settings. We can see that $\mE_1$, $\mE_2$, $\mE_3$ and $\mE_4$ have much bigger SEs than the Sine, Tent and Logistic maps in most parameter settings. Moreover, $\mE_3$ and $\mE_4$ have quite large SEs in the whole parameter settings that are close to the theoretical maximum value 10. With a larger SE, the outputs of a chaotic map distribute more random in the interval $(0,1)$. The average SEs of different chaotic maps in their respective chaotic ranges are listed in Table~\ref{table.averaRes}, from which we can also observe that the new chaotic maps have better ergodicity than their seed maps.

\subsection{Correlation Dimension}
The CD is a type of fractal dimensions and describes the dimensionality of the space occupied by a set of points. It is defined as Definition~\ref{defn.CD} and can be used to measure the strangeness of chaotic attractor.

\begin{table}[htbp]
\newcommand{\tabincell}[2]{\begin{tabular}{@{}#1@{}}#2\end{tabular}}
\renewcommand{\arraystretch}{1.3}
\setlength{\tabcolsep}{15pt}
\begin{center}
\caption{Average measure results of different chaotic maps within their respective chaotic ranges.}
\label{table.averaRes}
\begin{tabular}{|l|c|c|c|} \hline

  Chaotic maps &     SE & CD \\\hline\hline
  Sine         &  9.150352  & 0.904162 \\\hline
  $\mE_1$      &  \textbf{9.269728}  & \textbf{0.925160} \\\hline
  $\mE_2$      &  \textbf{9.662937}  & \textbf{1.613083} \\\hline\hline
  Logistic     &  9.072545  & 0.902477 \\\hline
  Sine         &  9.150352  & 0.904162 \\\hline
  Tent         &  7.389761  & 0.876820 \\\hline
  $\mE_3$      &  \textbf{9.663178}  & \textbf{1.103635} \\\hline
  $\mE_4$      &  \textbf{9.641178}  & \textbf{1.455635} \\\hline\hline

\end{tabular}
\end{center}
\end{table}

\begin{defn}
\label{defn.CD}
The CD of a time series $\{s_i~|~i=1,2,...,N\}$ is defined by
\begin{equation*}
  d = \lim_{r\rightarrow 0}\lim_{N\rightarrow\infty}\frac{\log C_e(r)}{\log r},
\end{equation*}
where $e$ is a given embedding dimension and $C_e(r)$ is the correlation integral which can be calculated as
\begin{equation*}
\begin{split}
   C_e(r) = &\lim_{N\rightarrow\infty}\frac{1}{[N-(e-1)\zeta][N-(e-1)\zeta-1]}\\
           & \times \sum_{i=1}^{N-(e-1)\zeta}~\sum_{j=i+1}^{N-(e-1)\zeta}\theta(r-|\bar{s}_i-\bar{s}_j|), \\
           %= &\lim_{N\rightarrow\infty}\frac{2}{N^2}\sum_{i=1}^N\sum_{n=1}^{N-i}\theta(r-|\bar{x}_n-\bar{x}_{i+n}|) \\
\end{split}
\end{equation*}
where $\theta(\omega)$ is a Heaviside step function, which is defined by
\begin{equation*}
  \theta(\omega) = \begin{cases} 0, \ \mbox{if}\ \omega \leq 0, \\
  1, \ \mbox{if} \ \omega > 0.\end{cases}
\end{equation*}
$\zeta$ is the time delay and is usually set to 1. The new data sequence $\{\bar{s}_t~|~t=1,2,3,...\}$ is
\begin{equation*}
\begin{split}
  \bar{s}_t &= (s_t,s_{t+\zeta},s_{t+2\zeta},...,s_{t+(e-1)\zeta}),\\
  t &= 1,2,...,N-(e-1)\zeta.
  \end{split}
\end{equation*}
\end{defn}

If it exists, $d$ is the slope of the log-log plot of $C_e(r)$ vs. $r$, which is defined by
\begin{equation*}
  d = \lim_{r\rightarrow 0}\lim_{N\rightarrow\infty}\frac{d[\log C_e(r)]/dr}{d(\log r)/dr}.
\end{equation*}

\begin{figure}[htbp]
\centering
\begin{minipage}[b]{0.98\linewidth}
  \begin{minipage}[b]{1\linewidth}
    \centerline{\includegraphics[width=1\linewidth]{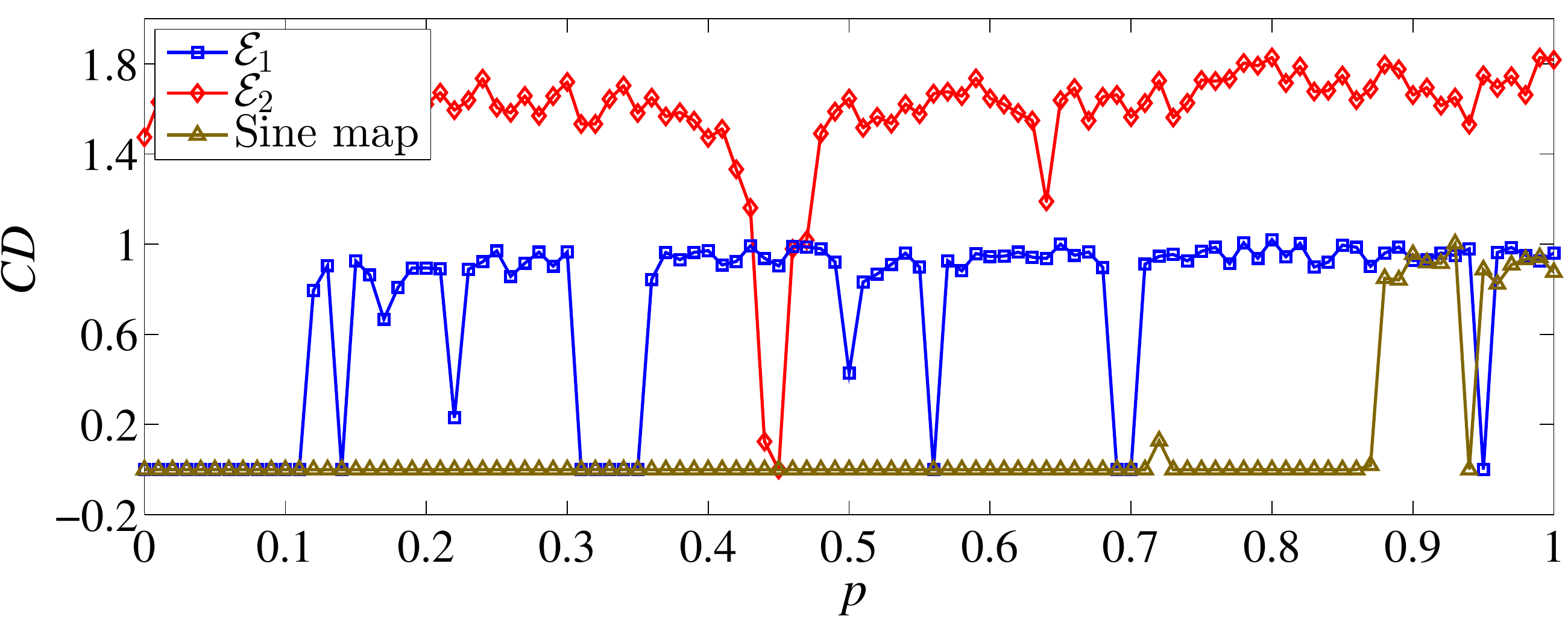}}
    \centerline{(a)}
  \end{minipage}\hfill
  \begin{minipage}[b]{1\linewidth}
    \centerline{\includegraphics[width=1\linewidth]{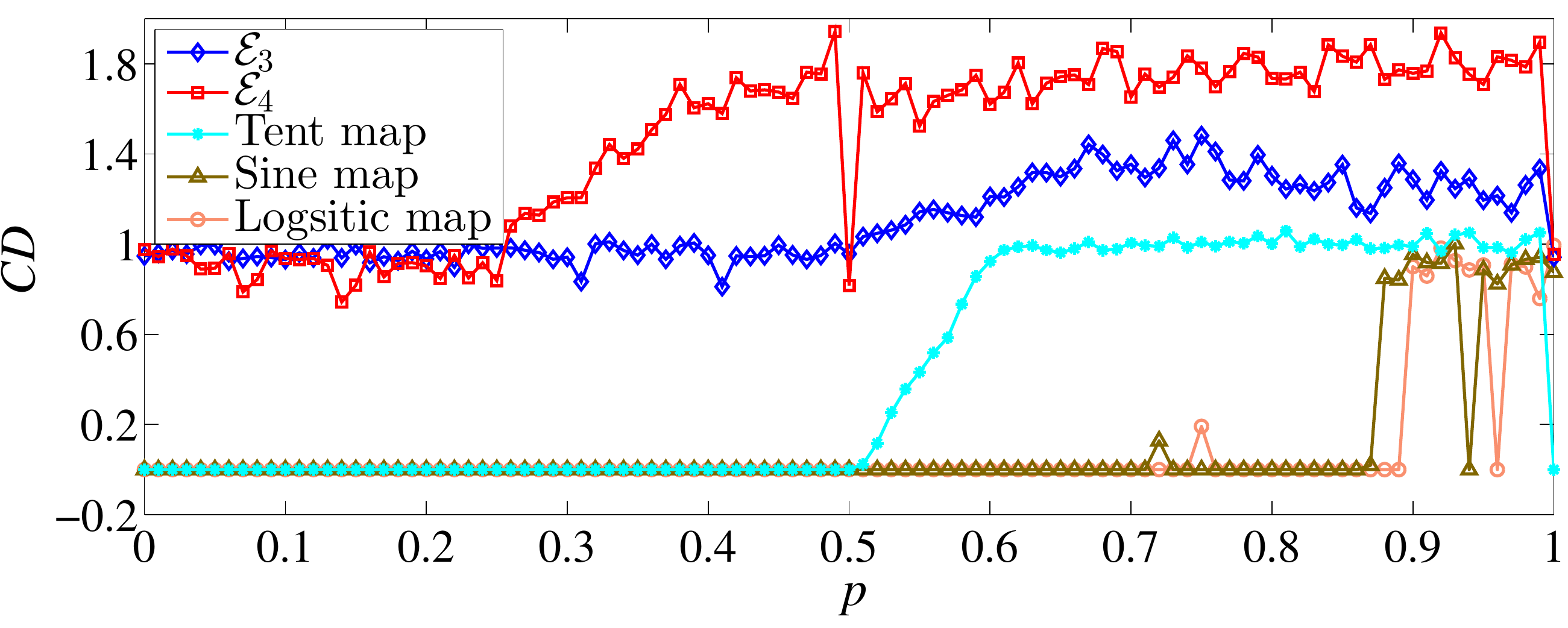}}
    \centerline{(b)}
  \end{minipage}\hfill
\end{minipage}\hfill
    \caption{(a) CD comparisons of $\mE_1$, $\mE_2$ and Sine map; (b) CD comparisons of $\mE_3$, $\mE_4$, Tent, Sine and Logistic maps.}
    \label{fig.CDs}
\end{figure}

The method proposed in~\cite{Albano1988PRA} is used to calculate the CDs of different chaotic maps and the embedding dimension $e$ is set as 2. Fig.~\ref{fig.CDs} plots the experimental results. As can be seen from the figures, the four new chaotic maps have much bigger CDs than their corresponding seed maps in most parameter settings. These seed maps have very small CDs that are close to 0 in many parameter settings. This means that their attractors have low degree of freedom. The average CDs of different chaotic maps in their respective chaotic ranges are listed in the third column of Table~\ref{table.averaRes}, in which we can get that $\mE_1$, $\mE_2$, $\mE_3$ and $\mE_4$ have larger CDs on average than their seed maps, which means that their attractors can occupy higher dimensionality in their phase planes to make their behaviors more irregular.

\subsection{Initial State Sensitivity}
The chaotic behavior has the property of initial state sensitivity, which means that an arbitrarily small change in the current state leads to significantly different future behaviors. The initial state sensitivity of a dynamical system can be measured by the correlation coefficient (CC), which is defined by Definition~\ref{defn.CC}.
\begin{defn}
\label{defn.CC}
The CC of two data sequences $\textbf{x}$ and $\textbf{y}$ is defined by
\begin{equation*}
  \label{correlation}
  CC(\textbf{x},\textbf{y}) = \frac{E[(\textbf{x}-\mu_{\textbf{x}})(\textbf{y}-\mu_{\textbf{y}})]}{\sigma_{\textbf{x}} \sigma_{\textbf{y}}},
\end{equation*}
where $\mu$, $\sigma$ and $E[.]$ denote the mean, standard deviation and expectation function, respectively.
\end{defn}
An absolute CC closing to 0 means that the two trajectories $\textbf{x}$ and $\textbf{y}$ have weak correlation.

\begin{figure}[htbp]
\centering
\begin{minipage}[b]{0.99\linewidth}
    \begin{minipage}[b]{.245\linewidth}
    \centerline{\includegraphics[width=1\linewidth]{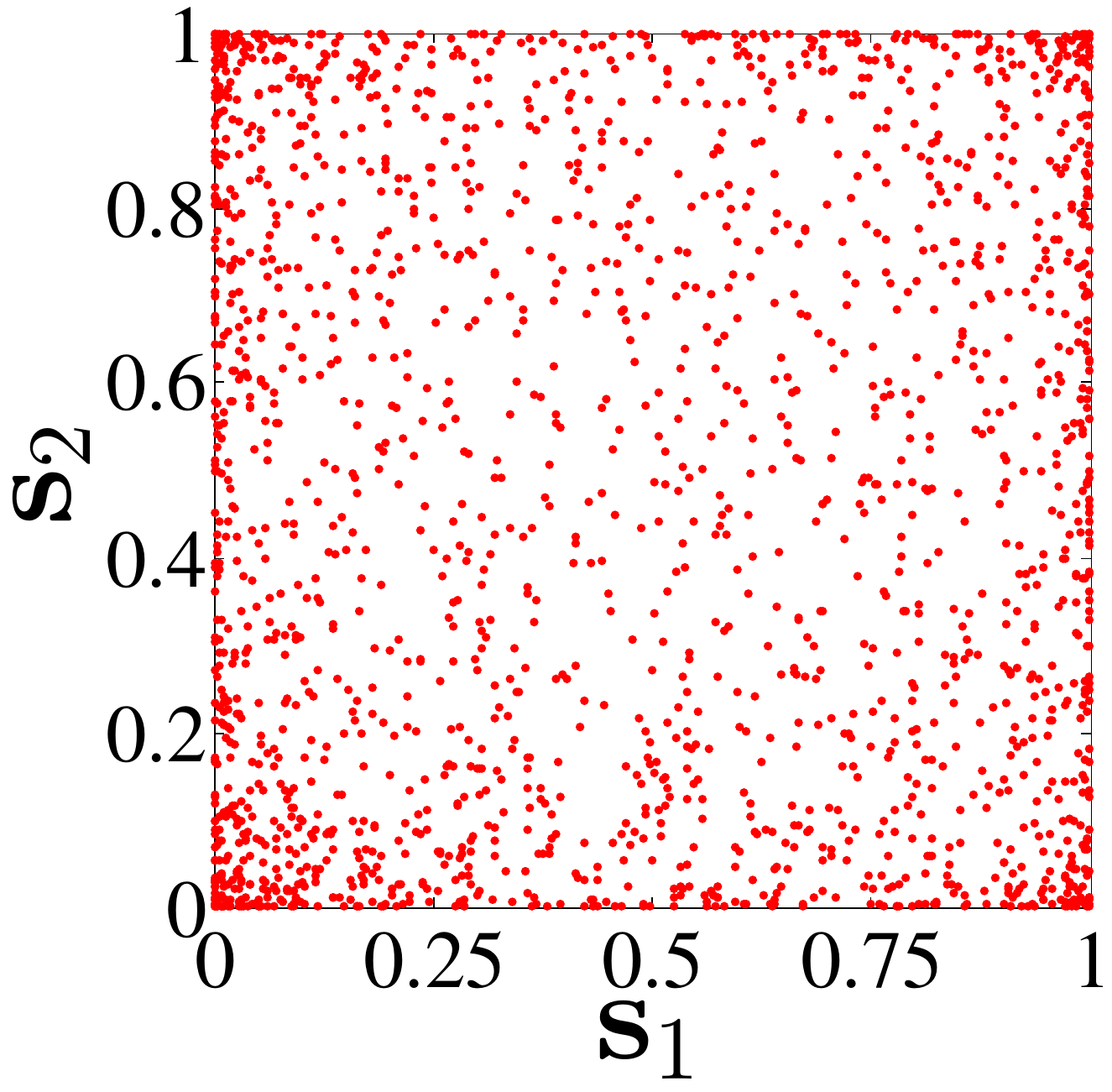}}
     \centerline{\includegraphics[width=1\linewidth]{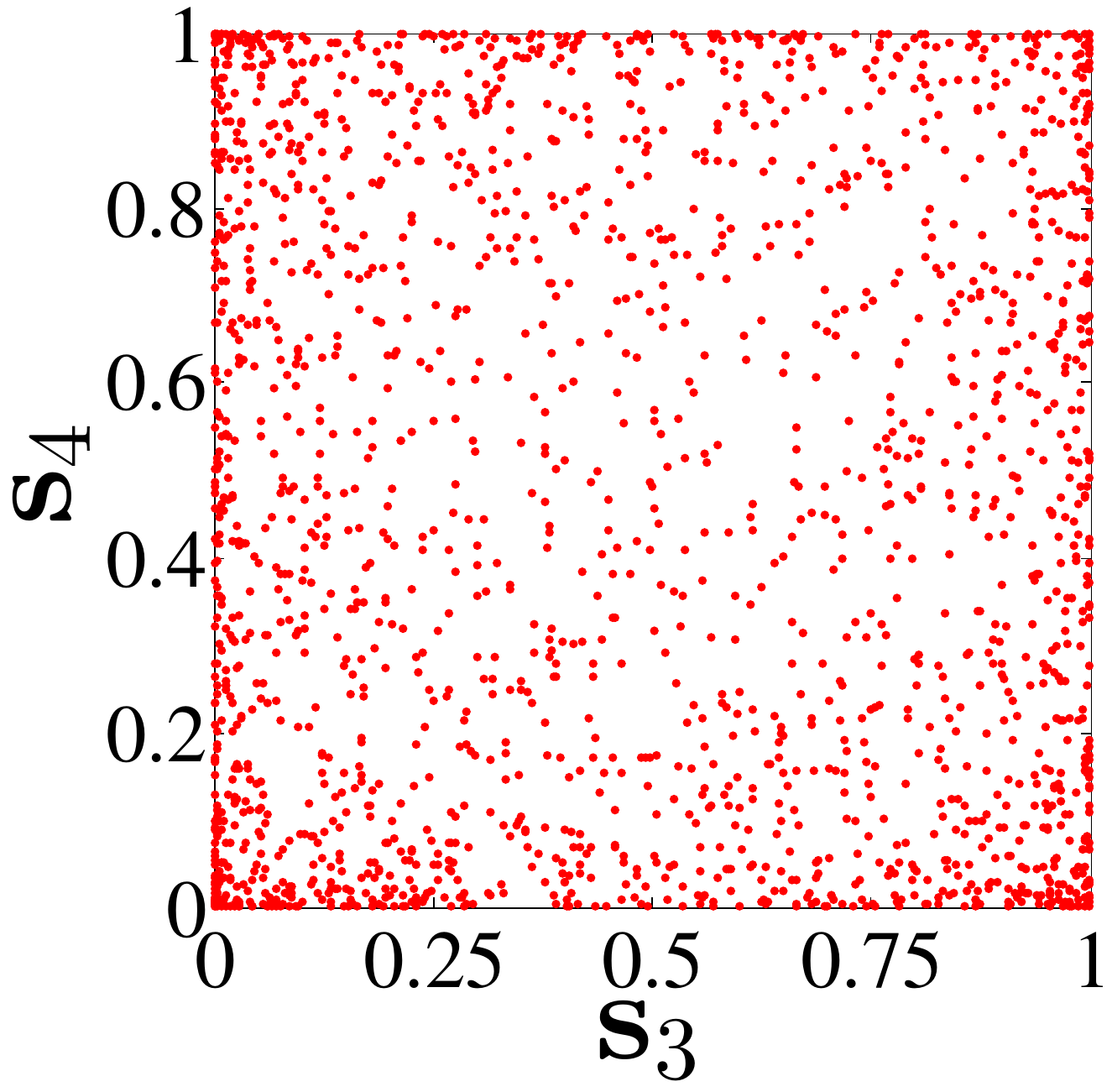}}
     \centerline{(a)}
  \end{minipage}\hfill
    \begin{minipage}[b]{.245\linewidth}
    \centerline{\includegraphics[width=1\linewidth]{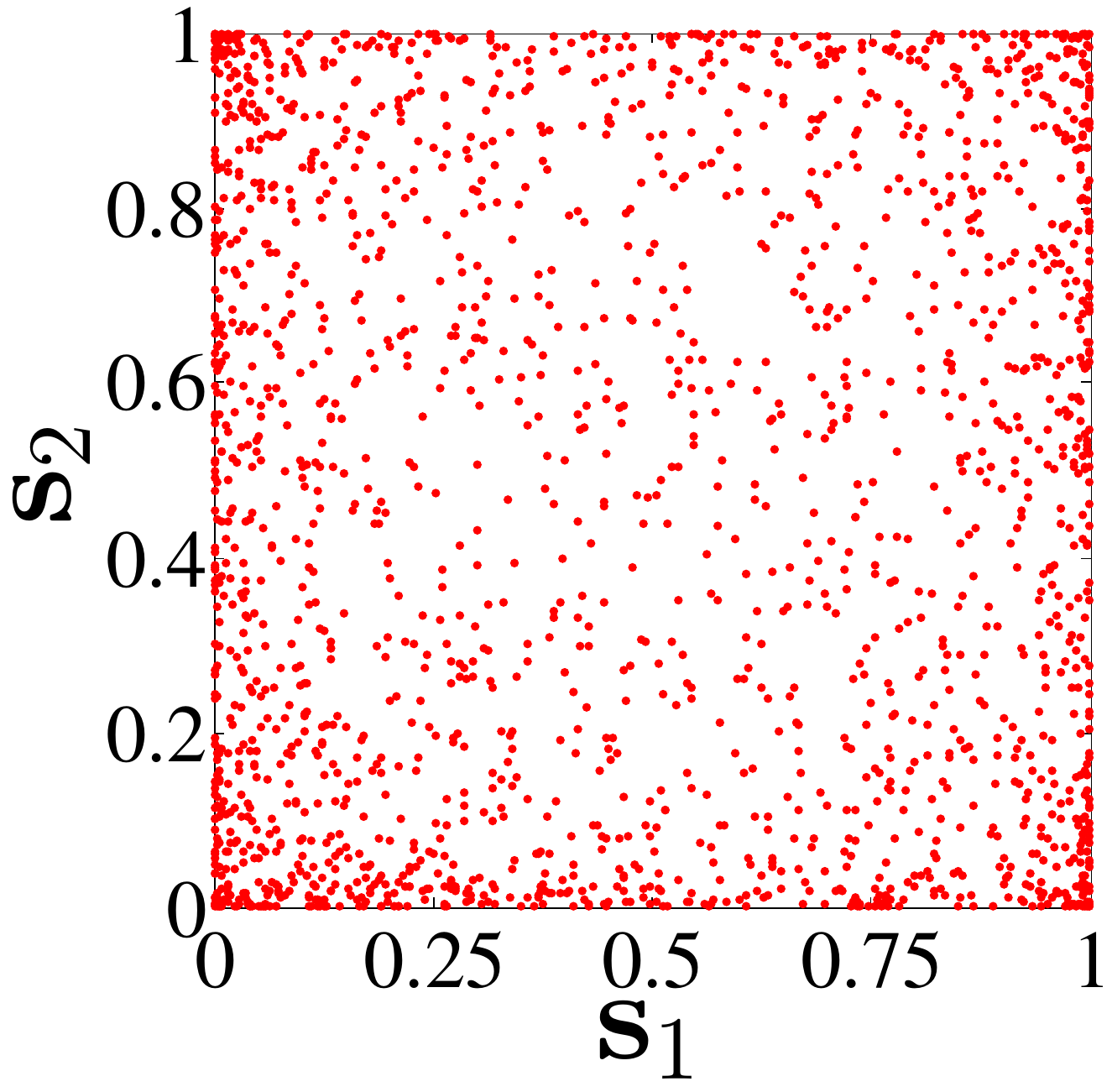}}
     \centerline{\includegraphics[width=1\linewidth]{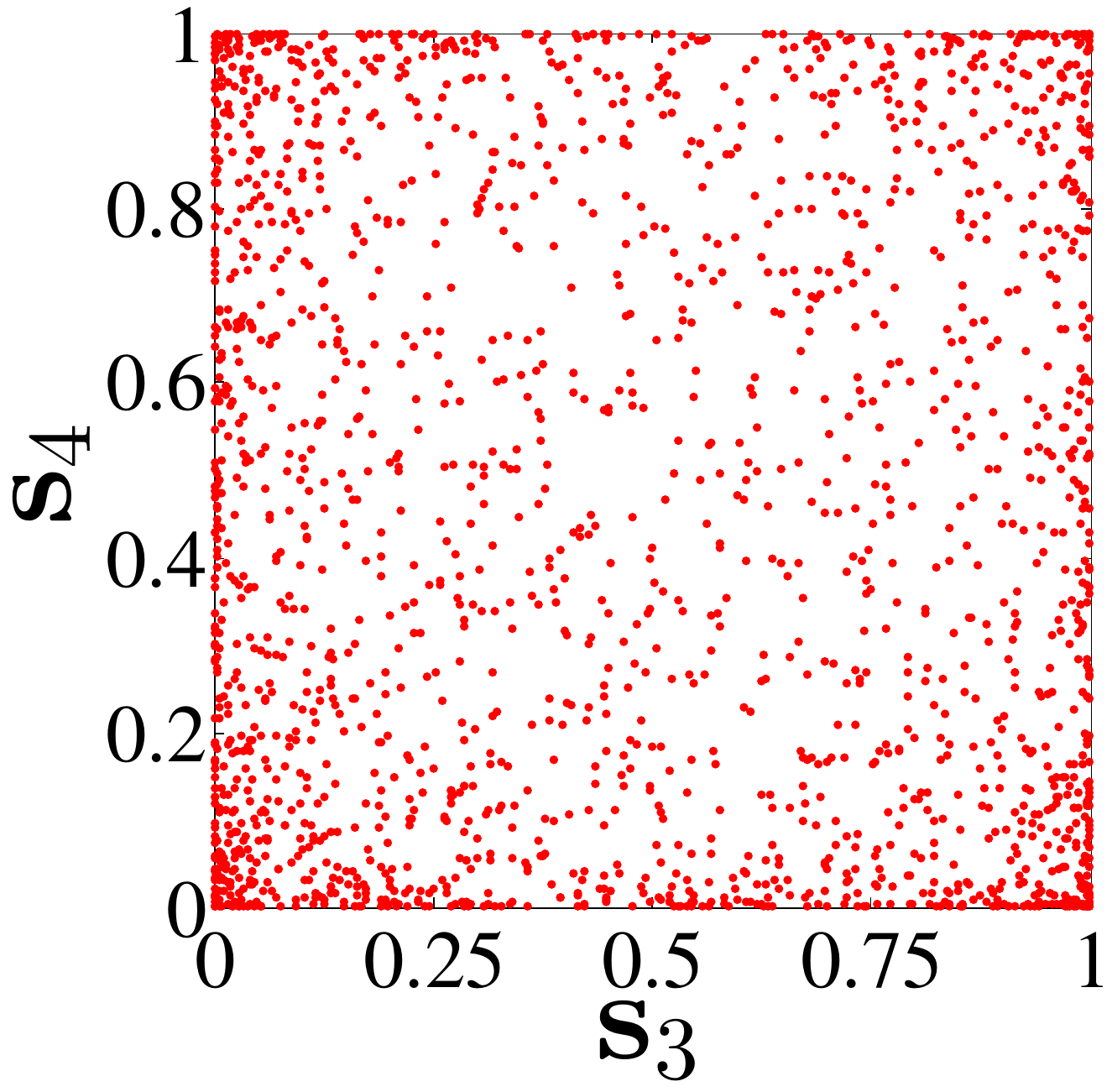}}
     \centerline{(b)}
  \end{minipage}\hfill
    \begin{minipage}[b]{.245\linewidth}
    \centerline{\includegraphics[width=1\linewidth]{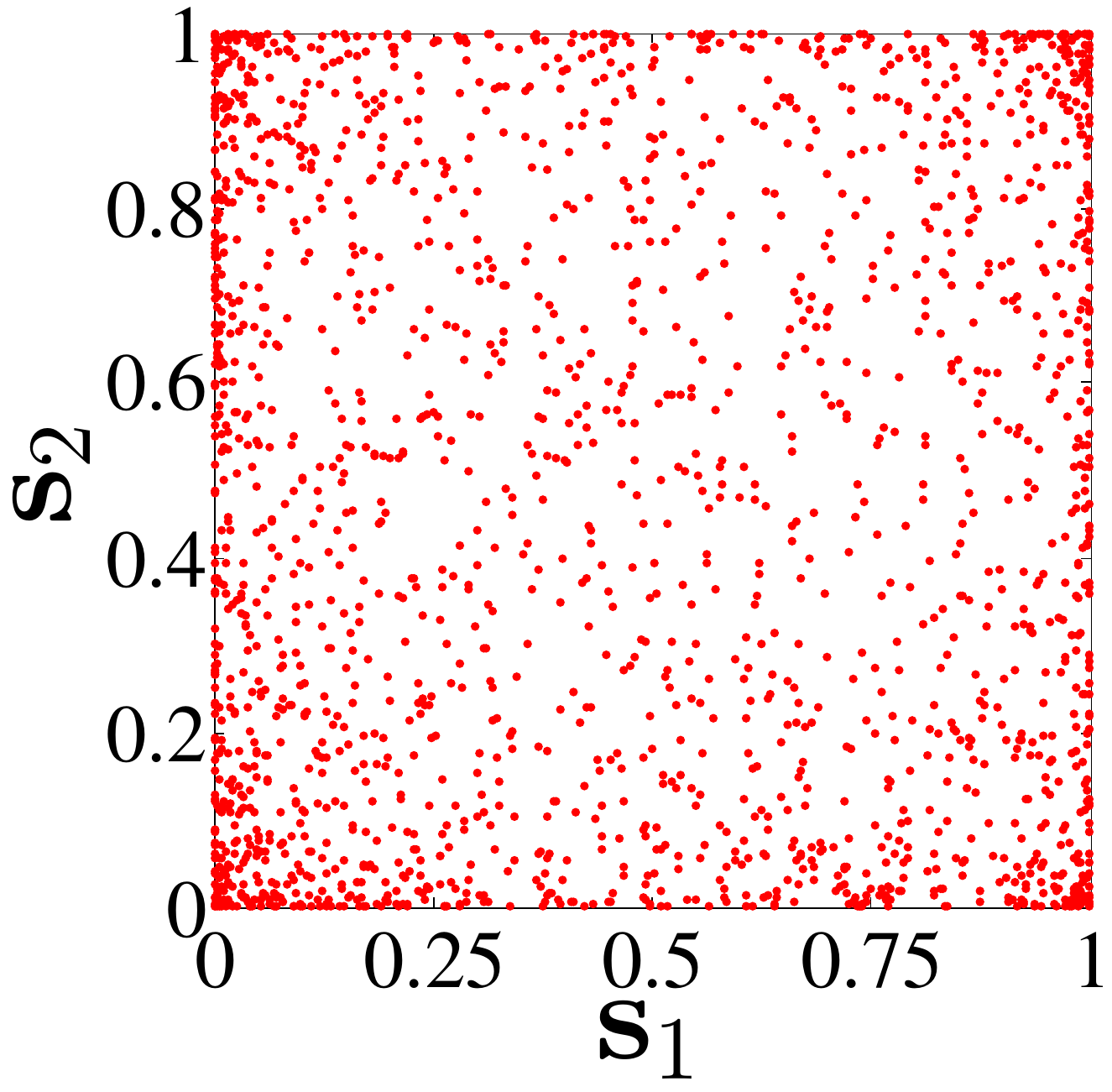}}
     \centerline{\includegraphics[width=1\linewidth]{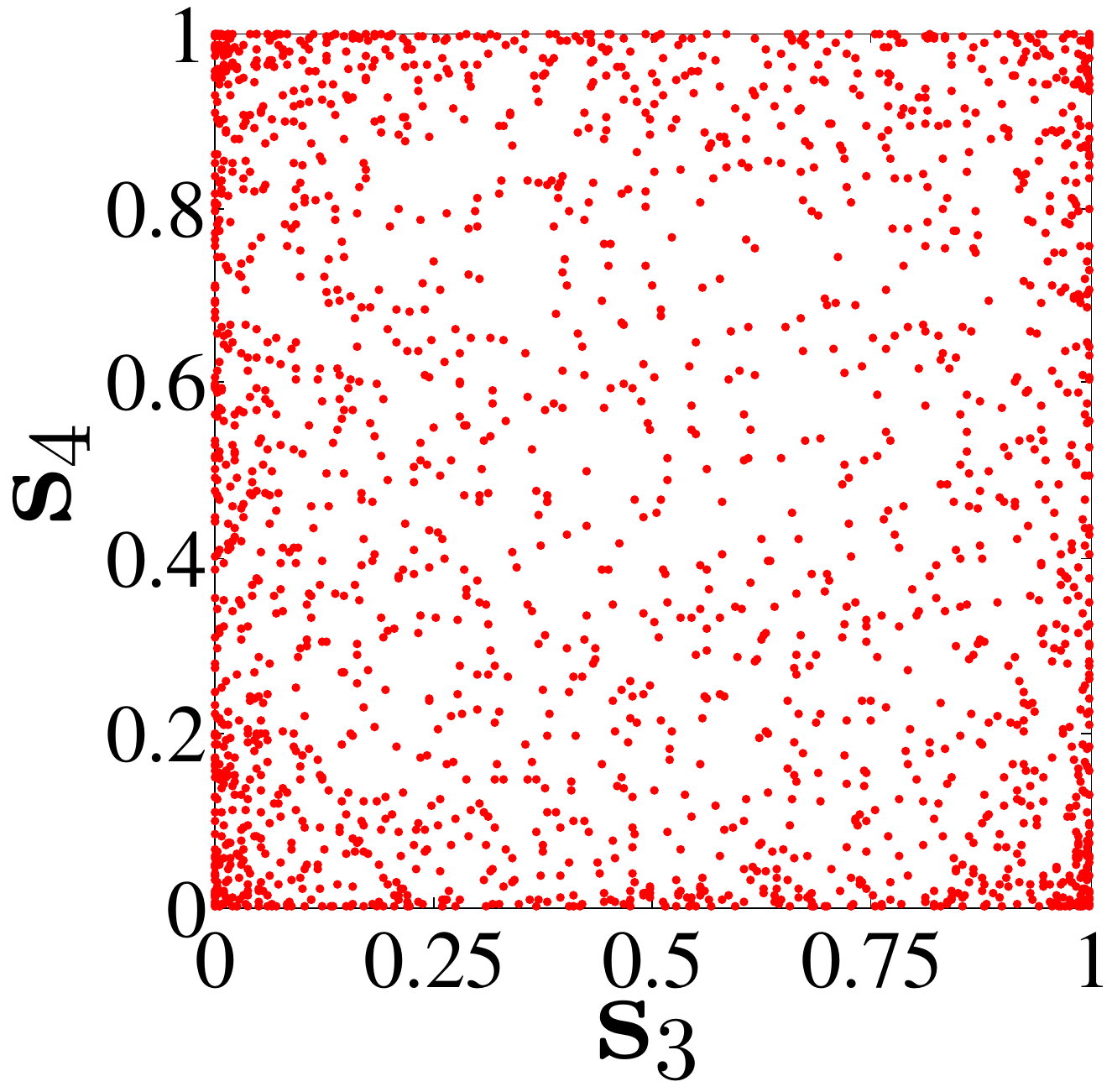}}
     \centerline{(c)}
  \end{minipage}\hfill
      \begin{minipage}[b]{.245\linewidth}
    \centerline{\includegraphics[width=1\linewidth]{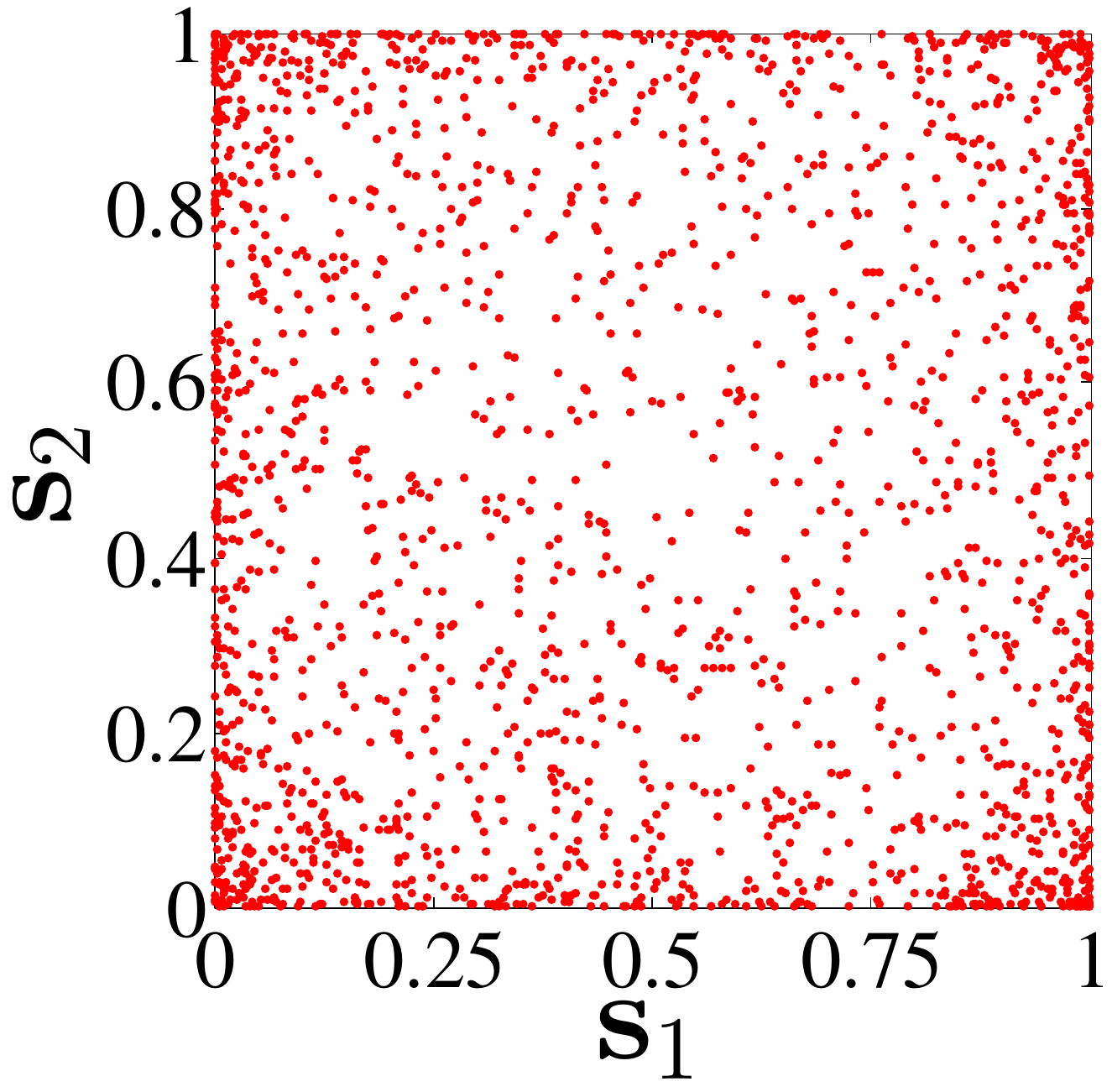}}
     \centerline{\includegraphics[width=1\linewidth]{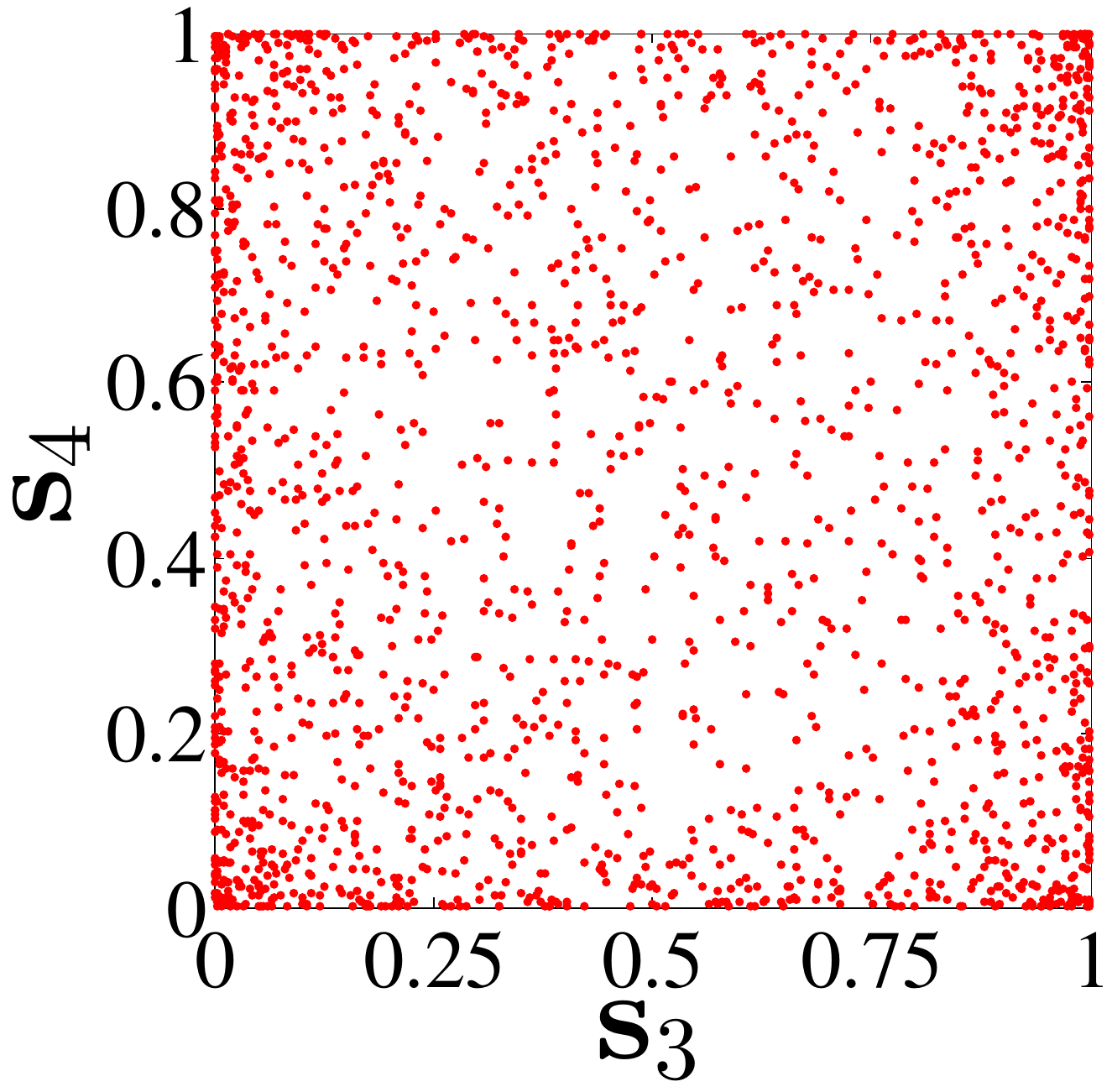}}
     \centerline{(d)}
  \end{minipage}\hfill
\end{minipage}\hfill
\caption{Output trajectories generated by (a) $\mE_1$, (b) $\mE_2$, (c) $\mE_3$, and (d) $\mE_4$ with a tiny change applied to their initial values (the top row) and control parameters (the bottom row).}\label{fig.Corre}
\end{figure}

For each chaotic map, the experiment was designed as follows: 1) apply a tiny change to the initial value and generate two trajectories $\textbf{s}_1$ and $\textbf{s}_2$ with the same control parameter; 2) apply to a tiny change to the control parameter and generate two trajectories $\textbf{s}_3$ and $\textbf{s}_4$ with the same initial value; 3) calculate CC between $\textbf{s}_1$ and $\textbf{s}_2$, and that between $\textbf{s}_3$ and $\textbf{s}_4$.
Table~\ref{table.corr} lists the average absolute CCs of different chaotic maps in their respective chaotic ranges. As can be seen from the table, the four new chaotic maps have much smaller absolute CCs on average than their seed maps, except for $\mE_2$ in applying a tiny change in initial value. Fig.~\ref{fig.Corre} plots the output pairs of ($\textbf{s}_1,\textbf{s}_2$) and ($\textbf{s}_3,\textbf{s}_4$) of these new chaotic maps. These output pairs randomly distribute in the whole phase plane, which straightforwardly display that they have weak correlations.
\begin{table}[htbp]
\newcommand{\tabincell}[2]{\begin{tabular}{@{}#1@{}}#2\end{tabular}}
\renewcommand{\arraystretch}{1.3}
\setlength{\tabcolsep}{15pt}
\begin{center}
\caption{Average absolute CCs of different chaotic maps within their respective chaotic ranges.}
\label{table.corr}
\begin{tabular}{|l|c|c|} \hline

  Chaotic maps&   CC($\textbf{s}_1$,$\textbf{s}_2$) &   CC($\textbf{s}_3$,$\textbf{s}_4$) \\\hline\hline
  Sine         & 0.173399 & 0.158888   \\\hline
  $\mE_1$      & \textbf{0.037798} & \textbf{0.029916}  \\\hline
  $\mE_2$      & \textbf{0.178791} & \textbf{0.032771}   \\\hline\hline
  Logistic     & 0.201365 & 0.204028  \\\hline
  Sine         & 0.173399 & 0.158888   \\\hline
  Tent         & 0.428486 & 0.413691  \\\hline
  $\mE_3$      & \textbf{0.020091} & \textbf{0.019763}  \\\hline
  $\mE_4$      & \textbf{0.098026} & \textbf{0.025788}  \\\hline\hline

\end{tabular}
\end{center}
\vspace{-5pt}
\end{table}
\section{Conclusion}
\label{section7}
This paper introduced the NCP model for generating new chaotic maps. It has six basic nonlinear operations, including the cascade, modulation, switching, fusion, scalar cascade and scalar modulation. Each operation is a general framework that uses existing chaotic maps as seed maps to generate new ones. The NCP model has excellent expansibility and can be extended by combining existing operations or introducing new nonlinear operations. The properties of the NCP model were discussed and its chaotic behavior was investigated using LE. Four examples of new chaotic maps were generated by the NCP model to show its effectiveness and their dynamics properties were carefully analyzed. Performance evaluations were performed in terms of LE, SE, CD and initial state sensitivity. Compared with existing ones, these newly generated chaotic maps have much wider chaotic ranges, their outputs are more random, their attractors have higher degree of freedom, and their initial states are more sensitive. Our future work will extend the NCP model by introducing new nonlinear operations.

\bibliographystyle{IEEEtran}
%\bibliography{ChaoticAri,secu_abrv}

\end{document}